\documentclass[twocolumn,showpacs,preprintnumbers,
amsmath,amssymb,aps,floatfix,pra,superscriptaddress]{revtex4}
\usepackage[dvips]{graphicx}
\usepackage{amsfonts}
\usepackage{dcolumn}
\usepackage{bm}
\usepackage{color}
\usepackage{epsfig}
\usepackage{hyperref}
\usepackage{bbold}

\newcommand{\ket}[1]{\lvert #1\rangle}
\newcommand{\bra}[1]{\langle#1\rvert}
\newcommand{\braket}[2]{\langle #1 \vert #2 \rangle}

\newcommand{\abs}[1]{\lvert#1\rvert}
\newcommand{\babs}[1]{\bigl\lvert#1\bigr\rvert}
\newcommand{\half}{\frac{1}{2}}
\newcommand{\HPtwoFone}{{}_2F_1}
\newcommand{\norm}[1]{\lVert#1\rVert}

\DeclareMathOperator{\sgn}{sgn}

\begin{document}

\def\jpb{J. Phys. B: At. Mol. Opt. Phys.~}
\def\pra{Phys. Rev. A~}
\def\prb{Phys. Rev. B~}
\def\prl{Phys. Rev. Lett.~}
\def\jmo{J. Mod. Opt.~}
\def\jetp{Sov. Phys. JETP~}
\def\etal{{\em et al.}}

\def\reff#1{(\ref{#1})}

\def\diff{\mathrm{d}}
\def\imagi{\mathrm{i}}

\def\beq{\begin{equation}}
\def\eeq{\end{equation}}

\def\expect#1{\langle #1 \rangle}

\def\vekt#1{\vec{#1}}
\def\vect#1{\vekt{#1}}
\def\vektr{\br}

\def\A{{\bf A}}
\def\J{{\bf J}}
\def\j{{\bf j}}
\def\q{{\bf q}}

\def\makered#1{{\color{red} #1}}

\def\tr{\mathrm{tr}}
\def\dspin{\hat{\bar{\psi}}}
\def\spin{\hat{\psi}}

\newcommand{\be}{\begin{equation}}
\newcommand{\ee}{\end{equation}}
\newcommand{\bea}{\begin{eqnarray}}
\newcommand{\eea}{\end{eqnarray}}
\newcommand{\br}{\mathbf{r}}
\newcommand{\e}{\mathrm{e}}

\def\fmathbox#1{\fbox{$\displaystyle #1$}}

\newcommand{\esssup}[1]{\mathop{\mathrm{ess}\,\sup}_{#1}}

\newtheorem{theorem}{Theorem}
\renewcommand{\thetheorem}{\hspace{-0.5em}}

\newtheorem{proof}{Proof}
\renewcommand{\thetheorem}{\hspace{-0.5em}}

\newtheorem{lemma}{Lemma}
\renewcommand{\thetheorem}{\hspace{-0.5em}}

\newtheorem{corollary}{Corollary}
\renewcommand{\thetheorem}{\hspace{-0.5em}}

\title{Density-potential mappings in quantum dynamics}

\author{M.~Ruggenthaler}
\affiliation{Department of Physics, Nanoscience Center, University of Jyv\"askyl\"a, 40014 Jyv\"askyl\"a, Finland}
\author{K.J.H.~Giesbertz}
\affiliation{Department of Physics, Nanoscience Center, University of Jyv\"askyl\"a, 40014 Jyv\"askyl\"a, Finland}
\affiliation{European Theoretical Spectroscopy Facility (ETSF)}
\author{M.~Penz}
\affiliation{Institut f\"ur Theoretische Physik, Universit\"at Innsbruck, 6020 Innsbruck, Austria}
\author{R.~van Leeuwen}
\affiliation{Department of Physics, Nanoscience Center, University of Jyv\"askyl\"a, 40014 Jyv\"askyl\"a, Finland}
\affiliation{European Theoretical Spectroscopy Facility (ETSF)}

\date{\today}

\begin{abstract}
In a recent letter [Europhys. Lett. {\bf 95}, 13001 (2011)] the question of whether the density of a time-dependent quantum system
determines its external potential was reformulated as a fixed point problem. This idea was used to generalize the existence and uniqueness
theorems underlying time-dependent density functional theory. In this work we extend this proof to allow for more
general norms and provide a numerical implementation of the fixed-point iteration scheme. 
We focus on the one-dimensional case as it allows for a more in-depth analysis using singular 
Sturm-Liouville theory and at the same time provides an easy visualization of
the numerical applications in space and time. We give an explicit relation between the boundary conditions on the density
and the convergence properties of the fixed-point procedure via the spectral properties of the associated Sturm-Liouville operator.
We show precisely under which conditions discrete and continuous spectra arise and give explicit examples. 
These conditions are then used to show that in the most physically relevant cases the fixed point procedure converges. 
This is further demonstrated with an example. 
\end{abstract}
\pacs{31.15.ee, 71.15.Mb, 31.10.+z }
\maketitle

\section{Introduction}

The essence of the many-body problem lies in our incapability of handling the huge number of degrees of freedom
of many-particle systems and consequently in our inability to determine the many-body states.
This problem spawned a lot of interest into the question whether one can devise a closed set of equations for
reduced quantities which do not involve the explicit solution of the Schr\"odinger equation and in which
the many-body correlations can be approximated efficiently. Pursuits in this direction have led to various
approaches such as many-body Green's function theory \cite{GF}, density matrix theory \cite{rDFMT, Klaas} and density-functional theory \cite{DFT, DFT2}.
These approaches differ in the complexity of the reduced quantity which is used to calculate the various observables
of interest. In this work we will focus on the simplest of these variables, namely the one-particle density, and 
ask the question to what extent this quantity determines the many-body states. 

Within the framework of time-dependent density functional theory (TDDFT) \cite{Peuckert, RG1984, CarstenBook} this question is asked for the special case 
that the density operator is linearly coupled to a scalar potential in the Hamiltonian. This linear coupling suggests
the possibility of a one-to-one relation between the scalar potential and the density and hence between densities and
wave functions. This fact was indeed proven by Runge and Gross \cite{RG1984} for the case that the potential has
a Taylor expansion in time and with the spatial boundary condition that the potential vanishes at 
infinity \cite{GrossKohn}. Another issue is whether a given density can be produced by some scalar potential. This
existence question, which is usually referred to as the $v$-representability problem is a more difficult one. 
The existence question is nevertheless an important one since it allows the construction of an effective noninteracting system
having the same density as the one of an interacting system and thereby convert the interacting problem into an
effective noninteracting one. This procedure is known as the Kohn-Sham method and forms the basis of virtually all
applications of TDDFT. The existence can be established under the condition that densities and potentials are Taylor
expandable in time \cite{RvL1999}.  This condition is sometimes too restrictive as has been discussed in, e.g. \cite{MaitraTodorov}. There are, however, indications that both the uniqueness and the existence theorems
of TDDFT are valid under more general conditions that do not require Taylor-expandability.
As a matter of fact, we know that Taylor-expandability is not a necessary condition for the validity of these theorems.
A first extension was given in \cite{RvL2001}  to the set of Laplace-transformable potentials under the assumption of a groundstate as an initial wave function. In \cite{RPB2010} a proof of the Runge-Gross theorem for dipole fields without restriction on the temporal form was presented. Recently, Tokatly in \cite{Tokatly2011} has given a rigorous proof for an arbitrary potential on a lattice. These findings demonstrate that the restriction to analytic potentials in time is not fundamental and we can extend the set of potentials beyond Taylor-expandable ones. 
Recently \cite{FP2011} we have introduced a new proof of the two basic theorems of TDDFT, i.e.\ the Runge-Gross theorem \cite{RG1984} and time-dependent v-representability theorem \cite{RvL1999}. We have reformulated the question whether a one-particle density is uniquely defined by an external potential in terms of a fixed-point problem. In this way we were able to lift the usual restriction of Taylor-expandable potentials and densities. 

Here we extend this proof to allow for 
norms on more general function spaces and provide additional mathematical details.
We focus on the one-dimensional case as it allows to use established mathematical methods
from singular Sturm-Liouville theory \cite{Zettl2005}. The Sturm-Liouville operator associated
with the density can be classified according to the boundary properties of the density
in which each class gives rise to specific spectral properties. 
We give an explicit relation between these spectral properties
and the convergence properties of the fixed-point procedure.
We show precisely under which conditions discrete and continuous spectra 
arise and give explicit examples. We finally provide a numerical implementation of the fixed-point 
iteration scheme for the case of periodic densities. 

The paper is organized as follows: In Sec.~\ref{SecTDDFT} we introduce the density-potential mapping and formulate the basic questions of this many-body theory as a fixed-point problem. We draw attention to the fundamental inequality that will give us the opportunity to derive uniqueness and existence of a fixed point. In Sec.~\ref{SecLinResp} we will derive in a general fashion the first part of the afore introduced inequality by using linear response theory. Then in Sec.~\ref{SecSturmLiouville} we will deduce the second part of the basic inequality by using Sturm-Liouville theory. With this we show in Sec.~\ref{secfixedpoint} uniqueness and existence of a fixed point. In Sec.\ref{SecCurrentDensity} we focus on periodic densities, derive the explicit form of the fixed-point iteration and show that an elementary numerical implementation of the proposed iteration converges. Finally we conclude in Sec.~\ref{SecCon}.

\section{Density-potential mapping as a fixed-point problem}

\label{SecTDDFT}

In this section we will introduce the fixed-point formulation of the density-potential mapping. We will formulate everything for simplicity in the one-dimensional case. Note, however, that the reasoning is independent of the dimension of the space in which the particles move. All considerations carry over to higher dimensional cases unchanged. 

The basic equation we want to examine is the non-relativistic equation of motion for a given initial state $\ket{\Psi(t_0)} = \ket{\Psi_0}$ of $N$ interacting particles, i.e.\ the time-dependent many-body Schr\"odinger equation (TDSE), 
\begin{eqnarray}
\label{TDSE}
\imagi \frac{\partial}{\partial t}  \ket{\Psi (t)} = \hat{H}([v],t) \ket{\Psi (t)}.
\end{eqnarray}   
The Hamiltonian in atomic units ($e = \hbar = m = 1$) is given by 
\begin{eqnarray*}
\hat{H}([v],t) =  \hat{T} + \hat{V}([v],t) + \hat{W} ,
\end{eqnarray*}
where the kinetic energy operator reads 
\begin{eqnarray*}
\hat{T} = \sum_{\sigma} \int  \diff x \; \hat{\psi}^{\dagger}(x \sigma) \left(- \frac{1}{2} \frac{\partial ^2}{\partial x^2}\right) \hat{\psi}(x \sigma),
\end{eqnarray*}
the interaction energy operator is 
\begin{multline*}
 \hat{W} = \frac{1}{2}\sum_{\sigma,\sigma'} \iint \diff  x \diff x' w (x-x') \hat{\psi}^{\dagger}(x \sigma) \hat{\psi}^{\dagger}(x' \sigma' ) \\
{} \times \hat{\psi}(x' \sigma') \hat{\psi}(x \sigma)
\end{multline*}
and 
\begin{eqnarray}
\label{ConjugateVariables}
 \hat{V}([v];t) = \int \diff x \,\hat{n}(x) v(x t)
 \end{eqnarray}
is the external energy operator with the density operator $\hat{n}(x) = \sum_{\sigma} \hat{\psi}^{\dagger}(x \sigma) \hat{\psi}(x \sigma)$. The operators $\hat{\psi}^{\dagger}(x \sigma)$ and $ \hat{\psi}(x \sigma)$ are the usual creation and annihilation field operators  for the spin $\sigma$ and $w (x-x')$ is the interaction potential. 

Keeping the initial state $\ket{\Psi_0}$ fixed for all further considerations, we observe that there is a mapping between external potentials $v(x t)$ and the time-dependent wave functions, i.e.\ each external potential generates an associated $\ket{\Psi([v],t)}$ by propagation of the Schr\"odinger equation. We assume that for every $v \in \mathfrak{V}$, where $\mathfrak{V}$ is the set of potentials under consideration, the Schr\"odinger equation (\ref{TDSE}) has a unique square-integrable solution $\ket{\Psi([v],t)}$. Actually, the wave functions are at least spatially two-times (weakly) differentiable \cite{BB2003}.

Since the wave functions are uniquely defined by the potential we immediately find that also all expectation values are uniquely determined by the potential, i.e.\ 
for a physical observable represented by an operator $\hat{O}$ we have
\begin{eqnarray*}
O([v],t) = \braket{\Psi([v],t)| \hat{O}}{\Psi([v],t)}.
\end{eqnarray*} 
The density is a special observable since it couples directly to the scalar potential in the Hamiltonian as is directly clear from 
Eq.~(\ref{ConjugateVariables}). Hence we may expect a one-to-one relation between densities  $n([v], x t)$ and potentials $v(x t)$.
However, since physical observables are gauge invariant such a one-to-one relation can only be expected up to a trivial
spatially constant shift $c(t)$ in the potential. This would mean that there are no two potentials differing more than a gauge 
that generate the same density. 
 If this is true, then the density uniquely determines the potential, i.e.\ $v([n], x t)$, and we find following the above reasoning, that the wave function 
 (up to a physically irrelevant phase factor) is uniquely determined by the density, i.e.\ $\ket{\Psi([n],t)}$. Consequently all observables become functionals of the density and we can in principle calculate all quantum mechanical expectation values by only knowing the density of the system. 

Our first task now is to determine a way to verify that the potential is determined by the density alone. To do so, a direct relation between both entities is desirable. The obvious way is to use the Schr\"odinger equation and deduce  such an interrelation. 
So we start by the evolution of the density, which is controlled by the Heisenberg equation of motion. This leads to the well-known continuity equation
\bea
\label{continuityeqn}
\partial_t n(x t) = - \partial_x j(x t),
\eea
where $\partial_x = \partial/\partial x$ and similarly for the time variable. Here $j(xt)$ is the expectation value of the current-density operator which is defined by
\be
\hat{j}(x) = \frac{1}{2 \imagi} \sum_{\sigma} \left[ \hat{\psi}^{\dagger}(x \sigma) \left( \partial_{x} \hat{\psi}(x \sigma)\right) - \left(\partial_{x} \hat{\psi}^{\dagger}(x \sigma)\right) \hat{\psi}(x \sigma)  \right]. \nonumber
\ee 
In order to make the dependence on the external potential explicit, we apply the Heisenberg equation on the current-density operator and find 
\begin{align}
\label{localforceeqn}
 \partial_t j(x t)& = - n(x t)  \partial_x v (x t)
- \bigl( \partial_x  T_{x x} (x t) + W_{x}(x t)  \bigr) , 
\end{align}
This equation describes the local-force density of the system. The momentum-stress tensor $T_{xx}(xt)$ is defined to be the
expectation value of the operator \cite{Martin1959}
\begin{multline*}
\hat{T}_{x x}(x) = \sum_{\sigma} \Bigl\{  \left(\partial _{x}  \hat{\psi}^{\dagger}(x \sigma ) \right) \partial_{x} \hat{\psi}(x \sigma) \\
 - \frac{1}{4}\partial^2_{x} \left(\hat{\psi}^{\dagger}(x \sigma)\hat{\psi}(x \sigma ) \right) \Bigr\},
\end{multline*}
and the divergence of the interaction-stress tensor $W_x (xt)$ is found as
the expectation value of
\begin{multline*}
\hat{W}_{x} (x) = \sum_{\sigma,\sigma'} \int \diff x'  \bigl(\partial_{x} w(x- x') \bigr) \hat{\psi}^{\dagger}(x \sigma) \hat{\psi}^{\dagger}(x' \sigma'  ) 
\\
{} \times \hat{\psi}(x' \sigma') \hat{\psi}(x \sigma).
\end{multline*}
By using the continuity equation (\ref{continuityeqn}) in Eq.~(\ref{localforceeqn}) we find an
explicit relation between the density and the potential:
\begin{eqnarray}
\label{InitialSturmLiouville}
 -\partial_{x}  \left[ n([v],x t) \partial_{x} v(x t) \right]   =   q([v], x t) - \partial^{2}_t n([v], x t) .
\end{eqnarray}
In this equation 
\begin{align*}
q([v], x t) &= \braket{\Psi([v],t)| \hat{q}(x)}{\Psi([v],t)},
\\
\hat{q}(x) &=  \partial_{x}\bigl(\partial_{x} \hat{T}_{x x}(x) +  \hat{W}_x (x)\bigr).
\end{align*}
Equation (\ref{InitialSturmLiouville}) enables us to investigate the density to potential mapping $n \mapsto v$.
We can do this by inserting a given density into the equation such that we obtain
\begin{eqnarray}
\label{InitialSturmLiouville2}
 -\partial_x   \left[ n(x t) \partial_x v(x t) \right]   =   q([v], x t) - \partial^{2}_t n(x t) .
\end{eqnarray}
We can then search for a potential $v$ that solves this equation. To do this we need also to give the initial state
in order to calculate $q([v],xt)$ on the right hand side of the equation.
The are now two cases to consider.
In the first case we assume that $n(xt)=n([u],xt)$, i.e.\ it is the density obtained by propagation of
the TDSE using some potential $u$ and the given initial state. We then know that $v=u$ is a solution to this equation.
If there is no other potential that solves the equation then there is a one-to-one correspondence between the
density and the potential. The Runge-Gross theorem is therefore equivalent to the uniqueness of a solution
of Eq.~(\ref{InitialSturmLiouville2}).
In the second case, we consider a density of which we do not a priori know whether it is obtained
from a time-propagation of the TDSE. In this case there are two possibilities to consider.
The first possibility is that there is no solution to  Eq.~(\ref{InitialSturmLiouville2}).
In that case the given density is not $v$-representable for the given initial state.
The second possibility is that we find a solution $v$. In that case it is not immediately clear
that this potential $v$ produces the given density. However, if we propagate the TDSE using this potential
we satisfy Eq.~(\ref{InitialSturmLiouville}) in which $q([v],xt)$ by construction is the same as in 
 Eq.~(\ref{InitialSturmLiouville2}). If we therefore subtract both equations we obtain
\be
\partial_t^2 \rho (x t) - \partial_x [ \rho (x t) \partial_x v (x t) ] = 0 \nonumber
\ee
for the density difference $\rho (x t) = n([v],x t)- n(x t)$.
For a given $v$ this a linear and homogeneous differential equation for $\rho (xt)$.
Let us now discuss its initial and boundary conditions. It follows immediately from the
equation of motion of the density operator that the density from the time-propagation satisfies
the conditions
\begin{align}
\label{eq:initdens}
\begin{split}
n(x t_0) &= \braket{\Psi_0| \hat{n}(x)}{\Psi_0}, \\
\left. \partial_t n(x t)\right|_{t_0} &= -\braket{\Psi_0|\partial_x \hat{j}(x)}{\Psi_0}.
\end{split}
\end{align}
Also the given density $n(xt)$ must satisfy these conditions otherwise we obviously can not find a potential $v$
producing this density and the chosen density would not be $v$-representable. If we therefore choose $n(xt)$ to have
these initial conditions then $\rho (xt)$ satisfies
\be
\rho (x t_0)=\partial_t \rho (x t_0)=0 . \nonumber
\ee
Furthermore to have a solution the density $n(xt)$ must have the same spatial boundary conditions as $n([v], xt)$ which
are dictated by the TDSE. This gives two further conditions. 
Finally we could add the additional condition that $\rho (xt)$ integrates to zero when integrating over space.
Since we have already five conditions on a linear differential equation that is second order in time and
first order in space it is clear the solution $\rho (xt) = 0$ is the only one. We thus find that 
if $v$ is a solution to Eq.~(\ref{InitialSturmLiouville2}) then $n(xt)=n ([v],xt)$.

After having fixed the boundary conditions for the potentials under consideration we can then ask the question whether there is any other 
solution $u \neq v$ generating the same density. This means we want to examine whether there is another 
$u \in \mathfrak{V}$ for which Eq.~(\ref{InitialSturmLiouville2}) holds such that by subtraction we would find
\begin{eqnarray}
\label{SturmLiouvilleDifference}
 - \partial_x   \left[ n(x t) \partial_x  \omega (xt)  \right]  =   q([v], x t) -  q([u], x t).
\end{eqnarray}
where we defined $\omega (xt) =  v(x t) -  u(x t)$.
In the case of the original Runge-Gross proof this question is answered by taking repeated time derivatives of this equation at the initial time $t_0$.
This assumes that all time derivatives of $n(xt)$ and $\omega (xt)$ in $t_0$ exist.
At time $t_0$ we have $q([v], x t_0)=q([u], x t_0) = \langle \Psi_0 | \hat{q} (x)| \Psi_0 \rangle$ and hence
\be
- \partial_x   \left[ n(x t_0) \partial_x   \omega (xt_0) \right]  =  0
\label{omega_zero}
\ee
Since $u$ and $v$ have the same boundary conditions $\omega (xt)$ vanishes at the boundaries and the unique solution is $\omega (xt_0)=0$.
If we now take the first time derivative of (\ref{SturmLiouvilleDifference}) we obtain the equation
\bea
\lefteqn{  - \partial_x   \left[ n(x t_0) \partial_x \partial_t \omega (xt_0)  \right] =} \nonumber \\
 && - \imagi \int dy \langle \Psi_0 | [\hat{q}(x), \hat{n} (y) ] | \Psi_0 \rangle \omega (yt_0) \nonumber \\
&& + \partial_x   \left[ \partial_t n(x t_0) \partial_x   \omega (xt_0) \right] \nonumber
\eea
Since the right hand side vanishes we find for $\partial_t \omega (xt_0)$ the same equation (\ref{omega_zero}) as for $\omega (xt_0)$
and we find that $\partial_t \omega (xt_0)=0$. Continuing this way we find that all time derivatives of $\omega (xt)$ vanish at
the initial time $t_0$, i.e.\ $\partial_t^k \omega (xt_0)=0$ for all integers $k \geq 0$.
We thus see that it is a necessary condition for two potentials to give the same density that all the time derivatives
of their difference in $t_0$ vanish. Therefore if one of those derivatives for some $k$ does not vanish the two potentials can not
give the same density.
However, it is still possible that all $\partial_t^k \omega (xt_0)=0$ while the potentials $u$ and $v$ are still different.
For example, a function of the form $\omega (xt) = f(x)\exp{(-1/(t - t_0)^2 )}$ has all its time-derivatives vanishing in $t_0$.
To eliminate such cases we have to demand that the function $\omega (xt)$ is equal to its Taylor expansion around $t_0$.
With this additional condition the vanishing of all $\partial_t^k \omega (xt_0)$  implies that $\omega (xt)=0$.
Therefore the mapping from the set of Taylor expandable potentials around $t_0$ with the given boundary conditions 
to the set of densities produced by it is one-to-one. This is the statement of the original Runge-Gross theorem.

\begin{figure}
\includegraphics[width=0.45\textwidth]{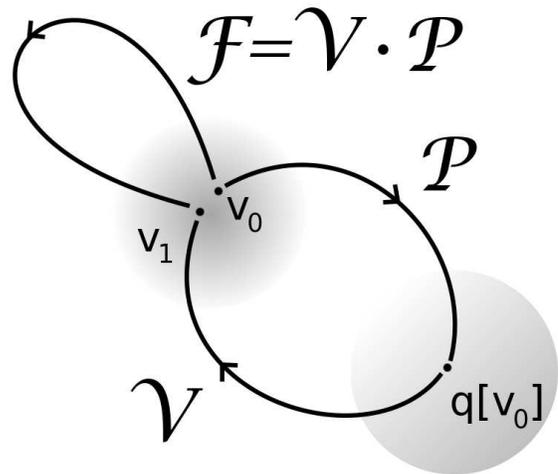} 
\caption{The potential-potential mapping $\mathcal{F}$ of Eq.~(\ref{mapF}) as composition of the mappings $\mathcal{P}$ and $\mathcal{V}$.}
\label{picture}
\end{figure}

Now one can pose the question, whether the Taylor-expandability of the external potentials is an essential condition for the Runge-Gross theorem to hold. To rephrase, can we prove the one-to-one mapping between the potentials and densities also for other, possibly more general sets $\mathfrak{V}$? As a matter of fact, we know this to be true. 
As discussed in the introduction there are already several extensions \cite{RvL2001,RPB2010,Tokatly2011}
which demonstrate that the restriction to analytic potentials in time is not fundamental. In this work we want to broaden the set $\mathfrak{V}$ of allowed potentials even further. We will do so by using Eq.~(\ref{InitialSturmLiouville2}) in order to define a mapping we will call $\mathcal{F}$, which maps potentials to potentials. The first part of this mapping concerns the right hand side of Eq.~(\ref{InitialSturmLiouville2}): we take a $v_0 \in \mathfrak{V}$ and propagate the initial state $\ket{\Psi_0}$ in a fixed finite time interval $[t_0,T]$ with this potential. From the associated wave function $\ket{\Psi([v_0],t)}$ we calculate $q([v_0], x t)$. This procedure we denote by 
\begin{equation*}
 \mathcal{P}: v_0 \mapsto q[v_0].
\end{equation*}
It calculates for every potential the corresponding divergence of the internal-force density. The second step of the mapping $\mathcal{F}$ we identify by the left hand side of Eq.~(\ref{InitialSturmLiouville2}): we take the previously determined $q([v_0], x t)$ and solve the linear differential equation (with the previously chosen boundary conditions)
\begin{eqnarray}
\label{InitialSturmLiouville3}
 -\partial_x    \left[ n(x t) \partial_x  v_1(x t) \right]   =   q([v_0], x t) - \partial^{2}_t n(x t), 
\end{eqnarray}
in order to calculate a new potential $v_1$. This operation we designate by
\begin{equation*}
 \mathcal{V} : q[v_0] \mapsto v_1.
\end{equation*}
It computes the potential to a given divergence of the local force-density and a chosen one-particle density $n$. Next we define the
combined map (see Fig.~\ref{picture}) 
\be
\mathcal{F}: v_0 \mapsto (\mathcal{V} \circ \mathcal{P}) [v_0] = v_1,
\label{mapF}
\ee
which transforms our original potential $v_0$ into $v_1$.
What is the connection between $\mathcal{F}$ and the original problem of Eq.~(\ref{InitialSturmLiouville2}) or synonymously Eq.~(\ref{SturmLiouvilleDifference})? 
There are now two cases to consider. In the first case we take the density in Eq.~(\ref{InitialSturmLiouville2}) to be one coming from a potential $v$, i.e.\ $n=n[v]$, then
clearly
\begin{eqnarray*}
\mathcal{F}[v] = v.
\end{eqnarray*} 
Hence, $\mathcal{F}$ maps solutions of Eq.~(\ref{InitialSturmLiouville2}) to itself. Thus solutions of the original equation are fixed points of the mapping $\mathcal{F}$. Therefore, instead of asking whether there is a unique solution to Eq.~(\ref{InitialSturmLiouville2})
 we can equivalently ask if $\mathcal{F}$ has a unique fixed point.

In the second case we can insert a density into Eq.~(\ref{InitialSturmLiouville2}) for which we do not a priori know whether it can be generated
by a potential $v$. Then the existence of a fixed point $v$ guarantees, as shown above, that the density is $v$-representable, i.e.\ $n=n[v]$.
Therefore existence of a fixed point is equivalent to the $v$-representability of the density.
The $v$-representability question is essential for the existence of a Kohn-Sham system in density functional theory, since there
we ask whether a given density obtained from an interacting system can also be reproduced in a noninteracting system.
Therefore the Kohn-Sham system only exists when the $v$-representability question can be answered positively.

A first trivial test of $v$-representability is that the given density $n(x t)$ has to fulfill the initial conditions, i.e.\ Eq.~(\ref{eq:initdens}). Hence, we have to have an appropriate initial state with the right density. Only then Eq.~(\ref{InitialSturmLiouville2}) can have a solution $v$ at all. In order to investigate existence of a solution of Eq.~(\ref{InitialSturmLiouville2}) we again have to fix a boundary condition. Assuming Taylor-expandability in time of both, the potential as well as of the density, it was shown in \cite{RvL1999} how to construct the unique potential. This proof of $v$-representability complemented the original Runge-Gross proof and lent justification to the time-dependent Kohn-Sham scheme. Here we want also to go beyond the assumptions of the original extension of the Runge-Gross proof and use a formulation in terms of a fixed-point problem.
\\\\
In order to answer the raised fixed-point questions we will apply the following inequality:
\begin{eqnarray}
\label{step3}
\| \mathcal{F} [v_1] - \mathcal{F} [v_0] \|_{\alpha} \leq a \|v_1 - v_0\|_{\alpha}
\end{eqnarray}
with $a < 1$ and where $\| \cdot \|_{\alpha}$ is an appropriate norm depending on a positive parameter $\alpha$ on the space of potentials. This inequality will directly guarantee uniqueness of a given fixed point and with some further work we can deduce under which conditions a fixed point exists. In order to derive inequality (\ref{step3}) we will first obtain an inequality of the form
\be
\| q [v_1] - q [v_0] \|_{\alpha} \leq \frac{C}{\sqrt[p]{\alpha}} \| v_1 - v_0 \|_{\alpha} ,
\label{step1}
\ee
where $C$ is a positive constant and where $p \geq 1$ is a constant determining the function space. We subsequently derive the inequality 
\be
\| \mathcal{F}[v_1] - \mathcal{F}[v_0] \|_{\alpha} \leq D \| q [v_1] - q [v_0] \|_{\alpha},
\label{step2}
\ee
for a positive constant $D$. Using these two inequalities (\ref{step1}) and (\ref{step2}) we can immediately construct the required Eq.~(\ref{step3}) where 
$a=CD/\sqrt[p]{\alpha}$. If we
choose the positive parameter $\alpha > (C D)^p$ then clearly $a <1$. We note here, that one could in principle also use different norms for the space of potentials and for the $q$-functions. However, for simplicity, we keep those spaces identical.

\section{General linear-response inequality and the $\alpha$-norm}

\label{SecLinResp}

The major ideas for inequality (\ref{step3}) and the subsequent fixed-point approach are found in the derivation of inequality (\ref{step1}). It will not only introduce the afore mentioned $\alpha$-norm, which is the most important ingredient for making the proof work, but due to its universality, the derivation allows for different explicit realizations. Thus we can formulate the derivation for a general operator $\hat{O}$ and can keep the associated function spaces undetermined for the moment being. Again, the derivation applies directly to the three-dimensional case. 
The general idea that we present is a very simple one. We just want to quantify the physically intuitive idea that if two potentials $v_0$ and $v_1$ are close
then also the expectation values $O[v_1]$ and $O[v_0]$ calculated from them by time-evolution of the TDSE are close in some norm.
We start by calculating the non-equilibrium linear response of an operator $\hat{O}$ with respect to a parameter $\lambda \in [0,1]$, i.e.
\begin{eqnarray*}
 \frac{\diff O([v_{\lambda}], x t)}{\diff \lambda}  = \lim_{\epsilon \rightarrow 0} \frac{O([v_{\lambda} + \epsilon \Delta v], x t)- O([v_{\lambda}], x t)}{\epsilon},
\end{eqnarray*}
where $v_{\lambda} = v_0 + \lambda \Delta v$ and $\Delta v = v_1-v_0$. A straightforward calculation in the interaction picture of quantum mechanics and expanding the associated evolution operator in powers of $\epsilon$ leads to
\begin{multline*}
\frac{\diff O([v_{\lambda}], x t)}{\diff \lambda}
 \\
= - \imagi \int_{t_0}^{t} \! \diff t' \int \! \diff x' \braket{\Psi_0| [\hat{O}_{H_{\lambda}}(x t), \hat{n}_{H_{\lambda}}(x' t')]}{\Psi_0} \Delta v(x' t').
\end{multline*}
where $[. \, ,.]$ is the usual commutator. The subindex $H_{\lambda}$ indicates the operators in the Heisenberg picture for the Hamiltonian $\hat{H}([v_{\lambda}], t)$, i.e.\ 
\be
\hat{O}_{H_{\lambda}}(t) = \hat{U}([v_{\lambda}]; t_0,t) \hat{O} \hat{U}([v_{\lambda}]; t,t_0) ,\nonumber
\ee 
with $\hat{U}([v_{\lambda}]; t,t_0)$ the unitary evolution operator associated with $\hat{H}([v_{\lambda}], t)$. 
Since the Hamiltonian $\hat{H}([v_{\lambda}], t)$ is explicitly time-dependent this evolution operator is a time-ordered exponential.
Then by the fundamental theorem of calculus \cite{BB2003} we arrive at
\begin{multline*}
O([v_1], x t)-O([v_0], x t)
= \int_0^1 \diff \lambda \, \frac{\diff O}{\diff \lambda} ([v_{\lambda}] , x t) 
\\
= \int_{t_0}^t \diff t' \int \diff x' \chi (x t, x' t') \bigl(v_1 (x' t') - v_0 (x' t')\bigr),
\end{multline*}
where we have defined 
\be
\chi (x t, x' t') = -i \,\int_0^1 \diff \lambda 
\braket{\Psi_0| [\hat{O}_{H_{\lambda}} (x t), \hat{n}_{H_{\lambda}}(x' t')]}{\Psi_0}. 
\label{linresp}
\ee
The linear response kernel $\chi$ is assumed to be bounded in some properly chosen function space with norm 
\be
\norm{ f(t) } = \left( \int \diff r |f(r t)|^{p} \right)^{1/p},
\label{pnorm}
\ee
with $p \geq 1$.
In the following we will always use this norm, unless explicitly stated otherwise.
It is now straightforward to derive that
\begin{eqnarray}
\|O([v_1],t)- O([v_0],t) \|^{p} \leq \tilde{C}^{p}(t) \int_{t_0}^t \| \Delta v(t') \|^{p},
\label{ineq1}
\end{eqnarray}
where the constant $\tilde{C}(t)$ is the operator norm defined as
\begin{align}
\tilde{C}^{p} (t) = \sup_{g \neq 0} \frac{\| (\chi g) (t) \|^p}{\int_{t_0}^t dt'  \| g(t') \|^{p} } .
\label{opnorm}
\end{align}
This constant has an intuitive interpretation; it simply compares the norm of $g$ to that of $\chi g$ and searches for its largest possible ratio, i.e.\
the maximum amplification.
We observe at this point, that $\chi = \chi[v_0, v_1]$ as can be seen directly from Eq.~(\ref{linresp}). Thus, the linear response kernel depends on the choice of $v_1$ and $v_0$. As a consequence also the operator norm $\tilde{C}(t) = \tilde{C}([v_0,v_1],t)$ has the same dependence. The integral on the right hand side of inequality (\ref{ineq1}) can now be manipulated as follows
\begin{multline}
\int_{t_0}^t dt' \, \| \Delta v(t') \|^p = \int_{t_0}^t dt' \, e^{-\alpha (t'-t_0)}  e^{\alpha (t'-t_0)}\| \Delta v(t') \|^p \\
\leq \|\Delta v \|_{\alpha,t}^p \int_{t_0}^t dt' e^{\alpha (t'-t_0)} \leq \| \Delta v \|_{\alpha,t}^p \frac{e^{\alpha (t-t_0)}}{\alpha} .
\label{ineqa} 
\end{multline}
In this equation we defined the norm \cite{esssup}
\begin{eqnarray*}
 \|\Delta v\|_{\alpha,t}^p = \sup_{t' \in [t_0,t]} \left( e^{-\alpha (t'-t_0)} \|\Delta v(t') \|^{p} \right).
\end{eqnarray*}
where $\alpha$ is an arbitrary positive number. Such norms are commonly used to prove existence of solutions
to differential \cite{Walter} and integral equations \cite{Bielecki,Light}. In Appendix \ref{AppendixNorms} we
show that all $\alpha$-norms are equivalent and hence we can change $\alpha$ without changing the function space
that we are considering.
We now insert inequality (\ref{ineqa}) into Eq.~(\ref{ineq1}), multiply both sides with $e^{-\alpha (t-t_0)}$ and take the supremum over 
$[t_0,t]$. We then obtain
\be
\| O[v_1]- O[v_0] \|^p_{\alpha,t} \leq \frac{C(t)^p}{\alpha}  \| \Delta v \|^p_{\alpha,t} , \nonumber
\ee
where $C(t) = \sup_{t' \in [t_0,t]} \tilde{C}(t')$. If we define
\be
\| f \|_{\alpha} = \| f \|_{\alpha,T}
\label{alphanorm}
\ee
and $C=C(T)$ we find
\be
\| O [v_1] - O [v_0] \|_{\alpha} \leq \frac{C}{\sqrt[p]{\alpha}} \| v_1 - v_0 \|_{\alpha} \nonumber
\ee
on $[t_0,T]$. In this derivation we did not  explicitly select a function space, i.e.\ picked a value for $p$. As we used the same norm for the potentials as well as the $O$-functions the choice of $p$ dictates the space of potentials. Thus the set of potentials under considerations $\mathfrak{V}$ is part of the space of functions $v$ which have finite $\norm{\cdot}_{\alpha}$-norm. 
If we now choose $\hat{O} = \hat{q}(x)$ then we have derived inequality (\ref{step1}) within the appropriate $\alpha$-norm.

\section{The $\mathcal{V}$-mapping}

\label{SecSturmLiouville}

In what follows we will derive the second inequality and then obtain the main statements of the fixed-point approach.
In order to get the inequality (\ref{step2})  we will use the norm (\ref{pnorm}) with $p=1$  for the case of periodic boundary conditions.
For more general cases we use the norm with $p=2$ since then we can make use of what is called Sturm-Liouville theory. 
However,  the inequality may well be valid for more general norms and hence for broader sets of potentials than treated here.
The restriction to one-dimensional Sturm-Liouville theory is convenient, as the theory and all its strongest statements are usually formulated for the one-dimensional case only \cite{Zettl2005}. For an extension to higher dimensions in the context of TDDFT the authors provided some results in \cite{MP2010,MP2011}.
\\
The  $\mathcal{V}$-mapping is defined by Eq.~(\ref{InitialSturmLiouville3}) and therefore by the solution $v_1$ to the
following inhomogeneous equation as a functional of its inhomogeneity $\zeta$
\begin{eqnarray}
\label{InitialSturmLiouville5}
 -\partial_x    \left[ n(x t) \partial_x  v_1 (x t) \right]   =   \zeta (xt), 
\end{eqnarray}
where
\be
\zeta (x t) = q([v_0], x t) - \partial^{2}_t n(x t).
\label{inhom}
\ee
We consider this equation on an interval from $a$ to $b$.
We already note that an important property of the function $\zeta$ is 
that it is orthogonal to the constant function, i.e.
\be
0 = \int_a^b \diff x \, \zeta (x t).
\label{zeroint}
\ee
This is a consequence of the fact that $q$ is a divergence and that the number of particles is conserved.
This fact will be important later in our discussion.
We can directly integrate the Eq.~(\ref{InitialSturmLiouville5}) and its general solution is given by
\be
v_1 (xt) = \int_a^b \diff y \, G_t (x,y) \zeta (y t) + c \int_a^x \diff y \frac{1}{n (y t)} + d,
\label{vsol}
\ee
in which
\be
G_t (x,y) = \frac{1}{2} \left[ \theta (y-x) - \theta (x-y) \right] \int_y^x \diff z \frac{1}{n (z t)}, \nonumber
\ee
where $\theta$ is the Heaviside step function and
$c$ and $d$ are constants determined by the boundary conditions. Note that the last two terms
in Eq.~(\ref{vsol}) simply represent the most general homogeneous solution (i.e.\ $\zeta=0$) of the differential 
equation (\ref{InitialSturmLiouville5}).
We see that in Eq.~(\ref{vsol}) the integrals may diverge when the density goes to zero at the boundaries.
Let us therefore first consider the case in which this does not happen. The only physical relevant case where
this applies is the case of periodic systems. In this case periodic boundaries are imposed on the Schr\"odinger 
equation and hence on the densities and potentials. We can then identify boundary $a$ with $b$ and we thus have a finite
domain on which we require $v_1(a)=v_1(b)$ as well as $v_1'(a)=v_1'(b)$, where the prime means a derivative with respect to the spatial coordinate. The solution (\ref{vsol}) with these boundary conditions is given by
\be
v_1 (xt) = \int_a^b \diff y \, K_t (x,y) \zeta (y t) ,
\label{vsol2}
\ee
where
\be
K_t (x,y) = G_t (x,y) - \frac{\eta (xt) \eta (yt)}{\int_a^b \diff y \frac{1}{n(yt)}} \nonumber
\ee
and
\be
\eta (xt) = \frac{1}{2} \left( \int_a^x \diff y \frac{1}{n (y t)} + \int_b^x \diff y \frac{1}{n (y t)}\right) .\nonumber
\ee
Then employing the periodicity of the density we can readily check that the kernel $K_t$ satisfies the boundary conditions
\begin{align*}
K_t (a,y) - K_t (b,y) &= 0,   \\ 
\partial_x K_t (a,y) - \partial_x K_t (b,y) &= \frac{1}{n(a t)}   
\end{align*}
and similarly in the $y$-variable.
Then as a consequence of these boundary conditions and the fact that $\zeta$ is orthogonal to the constant function (see Eq.~(\ref{zeroint}) ) we
see that $v_1$ satisfies the required boundary conditions.
From Eq.~(\ref{vsol2}) we now see that
\begin{align}
| v_1 (xt) |  &\leq  \int_a^b \diff y \, \abs{ K_t (x,y) } \abs{\zeta (y t) } \nonumber \\
 & \leq  \max_{y \in [a,b]} \abs{ K_t (x,y) } \int_a^b \diff y \abs{\zeta (y t) } .
 \label{vfinite}
\end{align}
Since $K_t$ is a continuous function on a finite domain it attains a maximum and we thus see that $v_1 (xt)$ is finite whenever
$|\zeta|$ is integrable. This implies in particular that $|v_1|$ is integrable itself and that
\be
\int_a^b \diff x \, \abs{ v_1 (xt) }  \leq D_t  \int_a^b \diff y \,\abs{\zeta (y t) }, \nonumber
\ee
where
\be
D_t = \max_{(x,y)} \abs{ K_t (x,y) }. \nonumber
\ee
Now we can use this inequality in our iteration scheme and consider
the distance between successive potentials of Eq.~(\ref{SturmLiouvilleDifference}) in the $p=1$ norm
of Eq.~(\ref{pnorm}):
\begin{align}
\| v_2 (t)-v_1 (t) \| = \int_a^b \diff x \, |v_2 (xt) - v_1 (xt) |\nonumber \\
\leq D_t \int_a^b \diff x \, |q ( [v_1] , xt) - q ([v_0] , xt) |  \nonumber
\end{align}
where we used the explicit form Eq.~(\ref{inhom}) of the inhomogeneity. From this inequality it immediately
follows that
\be
\norm{ v_2 -v_1  }_\alpha \leq D \norm{ q[v_1] - q[v_0]  }_\alpha \nonumber
\ee
where we used the $\alpha$-norm defined in Eq.~(\ref{alphanorm}) and
\be
D = \max_{t \in [t_0,T]} D_t . \nonumber
\ee
The result above was derived for non-vanishing densities. However, if the density vanishes at the boundaries then the integral in Eq.~(\ref{vsol})
may diverge. To treat this case we make use of singular Sturm-Liouville theory since this theory allows us to have divergent potentials
provided that they are square integrable. This then naturally leads to the consideration of the case $p=2$ in Eq.~(\ref{pnorm}).
We note that Eq.~(\ref{InitialSturmLiouville5}) has the form of a Sturm-Liouville boundary-value problem (see Appendix~\ref{AppendixSturmLiouville}) parametrically depending on $t$. One can always choose the boundary conditions in such a way that the Sturm-Liouville operator 
\be
\hat{S}_t = -\partial_x \left[n(x t) \partial_x\right] \nonumber
\ee 
is self-adjoint in the Hilbert space of square-integrable functions. Then we can solve Eq.~(\ref{InitialSturmLiouville3}) and thus properly define the mapping $\mathcal{V}$, since we are able to expand the inhomogeneity of Eq.~(\ref{InitialSturmLiouville3}) in terms of a time-dependent orthonormal eigenbasis $\{ \varphi_{i}(xt) \}$, i.e.
\be
 -\partial_x [n(x t) \partial_x  \varphi_i (xt) ] = \lambda_i (t)  \varphi_i (xt) . \nonumber
\ee
The eigenvalue $\lambda_0 (t)=0$ is a special one for which we can find the eigenfunction explicitly as
\be
\label{eigensolution0}
\varphi_0 (xt) = c_1(t)  + c_2(t) \int_{x_0}^x \diff y \, \frac{1}{n(yt)}, \nonumber
\ee
with $c_1(t)$ and $c_2(t)$ are constants and $x_0$ is an arbitrarily chosen point $a < x_0 <b$. The quantities $c_1(t)$ and $c_2 (t)$ 
are determined by the boundary conditions and the normalization. In the following we will always choose boundary conditions in such a way
that $c_2(t)=0$ and hence $\varphi_0 (xt)=c_1(t)$ is simply the constant function. The appearance of the constant function is a consequence of
the gauge freedom in the Schr\"odinger equation, i.e.\ a constant shift in the potential will not change the density.
For a more detailed discussion of the boundary conditions
we refer to Appendix~\ref{AppendixSturmLiouville}.
We can now expand $\zeta$ in the eigenfunctions as follows
\begin{eqnarray*}
\zeta (x t) = \sum_{i=1}^{\infty} \varphi_i(x t) \braket{\varphi_i(t)}{\zeta (t)} = \sum_{i=1}^{\infty} \zeta_i(t) \varphi_i(x t) ,
\end{eqnarray*} 
where we used the standard inner product
\be
\langle f | g \rangle = \int_a^b \diff x \, f^*(x) g(x) . \nonumber
\ee
The trivial zero eigenvalue does not appear in this expansion, since the scalar product of $\zeta$ with
the constant function is zero as noted before in Eq.~(\ref{zeroint}).
Therefore the solution to Eq.~(\ref{InitialSturmLiouville3})
\begin{eqnarray}
v_1(xt) = \sum_{i=1}^{\infty} \frac{\zeta_i(t)}{\lambda_i(t)} \varphi_i(x t) \nonumber
\end{eqnarray}
is perpendicular to the time-dependent constant function too. However, we can always add such a constant to the unique potential without changing the physics. 
For comparison we note that the solution can be written in a form analogous to Eq.~(\ref{vsol2})
\be
v_1 (xt) = \int_a^b \diff y \, \Gamma_t (x,y) \zeta (y t), \nonumber
\ee
where we defined the Green's function \cite{Greensfunctionnote}
\be
 \Gamma_t (x,y) = \sum_{i=1}^\infty \frac{1}{\lambda_i (t)} \varphi_i (xt) \varphi_i^* (yt) . \nonumber
\ee
Since $\lambda_0 = 0 < |\lambda_1| \leq |\lambda_2| \leq \dots$ we find that $v_1$ is square integrable if $\zeta$ is, because we have the simple inequality
\begin{multline*}
\norm{ v_1(t) }^2 = \sum_{i=1}^{\infty} \left\lvert \frac{\zeta_i(t)}{\lambda_i(t)} \right\rvert^2 
\leq \frac{1}{\lambda_1(t)^2} \sum_{i=1}^{\infty} \abs{\zeta_i(t)}^2  
\\
= \frac{1}{\lambda_1(t)^2} \norm{ \zeta (t)}^2 < \infty.
\end{multline*}
Thus the mapping $\mathcal{V}$ is well-defined if we assume $\zeta(t)$ to be square-integrable.
After we have shown that the mapping $\mathcal{V}$ is well-defined, we can in a next step use the expansion in an eigenbasis to derive the second inequality. If we look at two successive potentials, say $\mathcal{F}[v_1] = v_2$ and $\mathcal{F}[v_0] = v_1$, we have in accordance to Eq.~(\ref{SturmLiouvilleDifference})
\begin{align*}
 - \partial_x   \bigl[ n(r t) \partial_x   \bigl( v_2(x t) - v_1(x t) \bigr) \bigr]   =   q([v_1], x t) - q([v_0], x t).
\end{align*}
We can then expand $v_2-v_1$  and $q[v_1]-q[v_0]$ in terms of the eigenfunctions of $\hat{S}_{t}$
and similarly obtain
\begin{eqnarray}
\label{EigenfctEstimation}
\| v_2 (t) - v_1 (t) \| \leq \frac{1}{| \lambda_1(t) |} \| q[v_1](t) - q[v_0](t) \|. 
\end{eqnarray}
If we now multiply Eq.~(\ref{EigenfctEstimation}) with $e^{-\alpha (t-t_0)}$ and take the supremum over the interval $[t_0,T]$ we arrive at
inequality (\ref{step2}) in which 
\be
D = \max_{t \in [t_0,T]}\left\{ \frac{1}{ | \lambda_1(t) | } \right\} . \nonumber
\ee
%
%
As pointed out before, in general we find a self-adjoint operator $\hat{S}_t$ with well-known spectral properties. 
These spectral properties can be related to the behavior of the density close to the boundary and a full classification of
the different cases can be found in the Appendix~\ref{AppendixContinuum}.
We either will have a discrete spectrum of normalizable eigenfunctions or we have a continuous spectrum 
of generalized (non-normalizable) eigenfunctions or a combination of both.
In all cases we can expand in those eigenfunctions since
in the case that we have a continuous spectrum we can replace the sum over eigenvalues in the expansions by an integral (see Appendix~\ref{AppendixInversion}). 
However, to derive our inequality we then have to make sure that there is a spectral gap between the zero eigenvalue and the continuum. We give an explicit example where we deduce a continuous spectrum gapped away from zero and the associated generalized eigenfunctions in Appendix~\ref{ap:ContinuumGap}. In Appendix~\ref{AppendixContinuum} we show that a spectral gap exists whenever the density does not decrease faster than a quadratic function to zero at the boundaries.

\begin{figure}[t]
\includegraphics[width=\columnwidth]{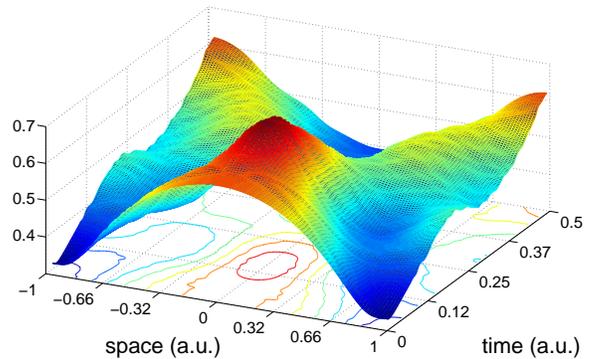} 
\caption{(color online). The density evolving in time from $t=0$ until $T=0.5$.}
\label{fig:density}
\end{figure}

\begin{figure}[b]
\includegraphics[width=\columnwidth]{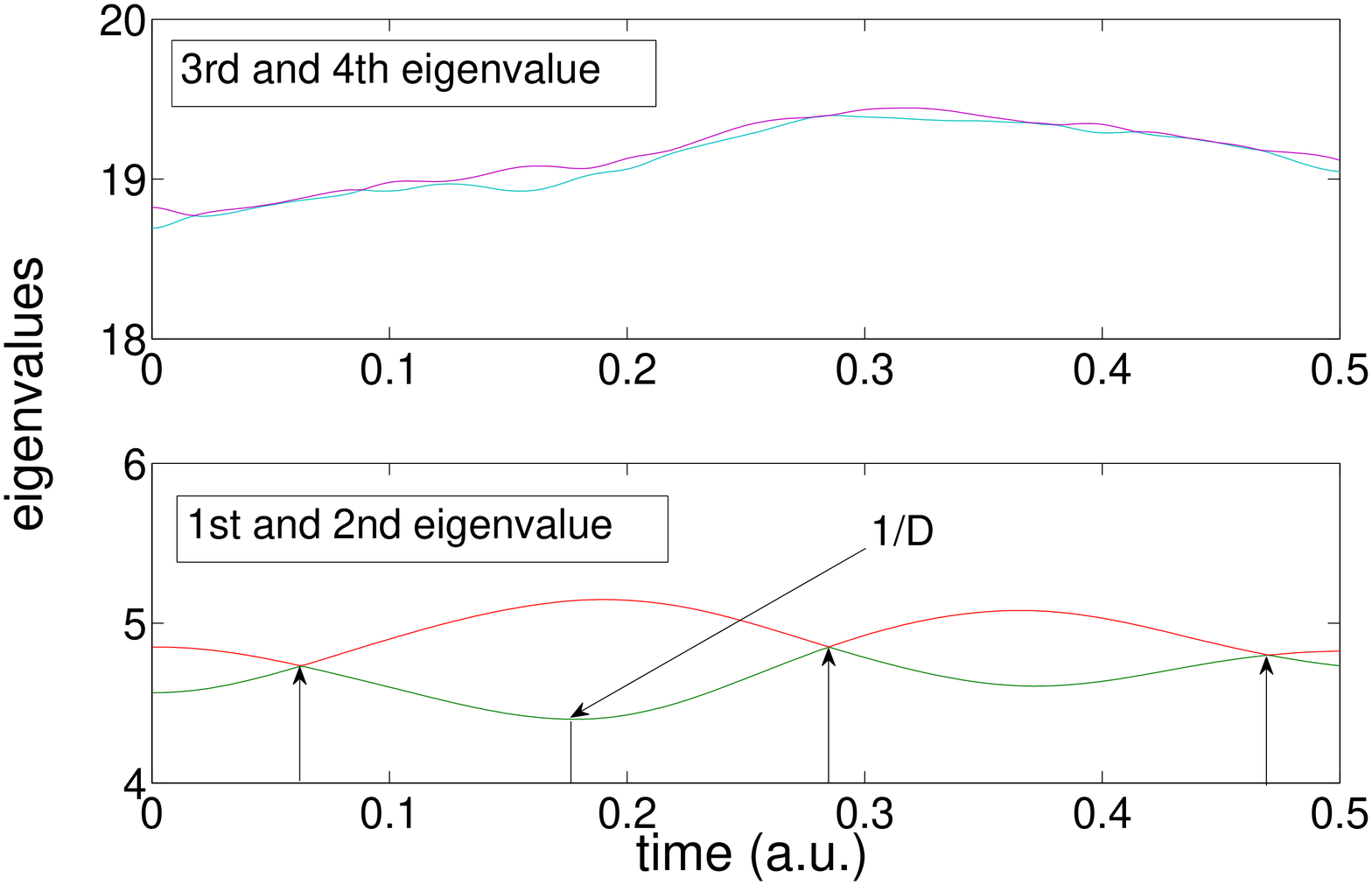} 
\caption{(color online). The four lowest eigenvalues in time from $t=0$ until $T=0.5$ for the Sturm-Liouville eigenvalue problem. The arrows indicate the time of degeneracy of the two lowest lying eigenfunctions. The constant $1/D \simeq 4.39$ as indicated at time $t \simeq 0.17$}
\label{fig:eigenvalues}
\includegraphics[width=\columnwidth]{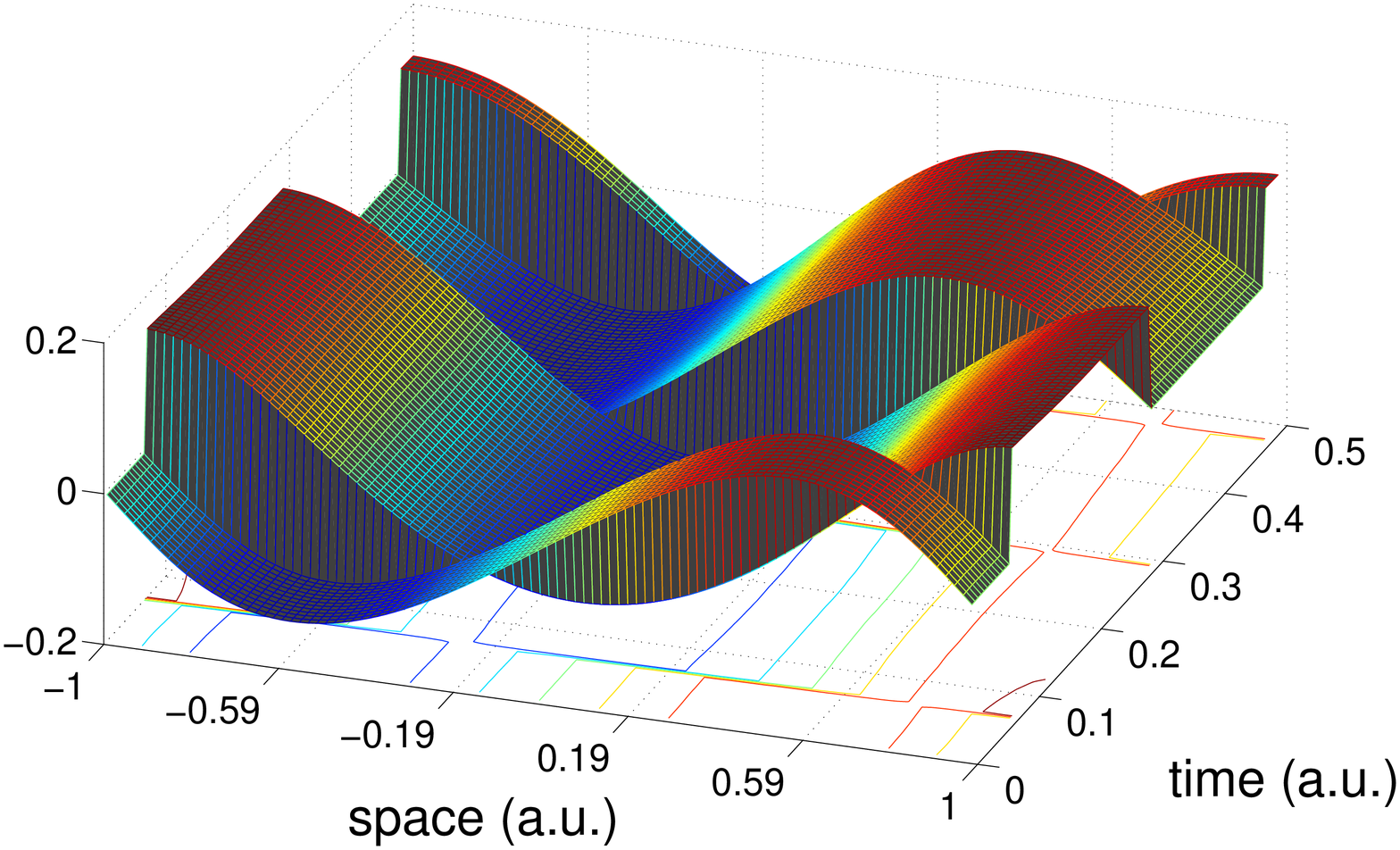} 
\caption{(color online). The lowest eigenfunction in time from $t=0$ until $T=0.5$ for the Sturm-Liouville eigenvalue problem. The lowest lying eigenfunction at certain points becomes degenerate with the second lying eigenstate, whereafter they change position in the spectrum.}
\label{fig:eigenstate1}
\end{figure}

Let us now give an explicit example of a Sturm-Liouville equation for which we calculate
the eigenvalues and eigenfunctions in time. The example involves the physical situation in which we have a single
particle on a ring, i.e.\  the Schr\"odinger equation on the interval from $-1$ to $1$ with periodic boundary conditions (the point $1$ is identified
with $-1$). As initial state we choose 
\be
\Psi_0(x) = c \left(  \exp(-1/(1-x^2))+1\right),
\label{initial}
\ee
where $c$ is the normalization constant that normalizes the wave function to one. 
We propagate this initial state with the external potential 
\be
v(x t) = \sin^2 (\pi x) \sin(10 \,t)
\label{potexample}
\ee
in atomic units for a short period of time, say from $t_0=0$ to the final time $T=0.5$ and calculate the density $n(xt)$ and $\hat{S}_t$. 
We use a Crank-Nicholson scheme for time-propagation on an equidistant grid and calculate the time-dependent density (see Fig.~(\ref{fig:density})). By diagonalizing the Sturm-Liouville operator with periodic boundary conditions we find the eigenvalues as well as eigenfunctions in time (see Fig.~(\ref{fig:eigenvalues}) and (\ref{fig:eigenstate1})). From this we deduce the value of the constant in inequality (\ref{step2}) for the current example as $D = \max_{t \in [t_0,T]}\{ |\lambda_1(t)|^{-1}\} \equiv 1/\min_{t \in [t_0,T]} \{|\lambda_1(t)|\} \simeq 0.23$. We note that at certain times the eigenvalues get degenerate (indicated with an arrow in Fig.~(\ref{fig:eigenvalues})) and 
therefore at the crossing point the eigenfunction with the lowest eigenvalue changes discontinuously
without changing the number of nodes. This is clearly visible in Fig.~(\ref{fig:eigenstate1}).
For the lowest lying non-trivial eigenfunctions (in the periodic case this amounts to two nodes) we can see how they change at the indicated times. 
This is a special feature of the periodic case, which cannot happen in the case that the boundary condition at one boundary does not depend
on the boundary condition at the other boundary. 
The calculated eigenvalues (see Tab.~(\ref{Table})) and eigenfunctions nicely agree with calculations done using the SLEIGN2 Sturm-Liouville code \cite{Bailey2001}.

\begin{table}[t]
\caption{The four lowest eigenvalues of the Sturm-Liouville eigenvalue problem for $t_0=0$ and $T=0.5$ in atomic units \label{Table}}
\begin{ruledtabular}
 \begin{tabular}{l d d}
  & \multicolumn{1}{r}{$t_0 = 0$} & \multicolumn{1}{r}{$T=0.5 $} \\ 
\hline 
 1st & 4.5646 & 4.7333 
\\
2nd & 4.8502 & 4.8252
\\
3rd& 18.6924 & 19.0458
\\
4th& 18.8226 & 19.1178
\\
\end{tabular}
\end{ruledtabular}
\end{table}
Before we use the derived inequality (\ref{step3}) to deduce uniqueness and existence of a fixed point we will shortly summarize our assumptions. 
For the periodic case where the density is non-vanishing we work with the $p=1$ norm and hence we require $q([v], x t)-\partial_t^2n(x t)$ 
to be integrable. This automatically implies via inequality (\ref{vfinite}) that the potential is finite everywhere.
For the case, of vanishing densities we use Sturm-Liouville theory and the $p=2$ norm, and we therefore
demand $q([v], x t)-\partial_t^2n(x t)$ to be square-integrable in the domain of $\hat{S}_t$, i.e.\ those $v$ that fulfill the chosen boundary and some further regularity conditions (they are in $\text{D}_{\text{max}}$ as defined in Appendix~\ref{AppendixSturmLiouville}).
In fact, we can weaken the requirements somewhat, since for the inequalities (\ref{step1}) and (\ref{step2}) we only need the above mentioned properties
for the difference potential $v_2 -v_1$ as well $q [v_1] -q[v_0]$, meaning that the potentials at the various iterations may have singularities provided that they occur
at the same spatial points. 

\section{Uniqueness and existence of a fixed point}

\label{secfixedpoint}

\begin{figure}[b]
\includegraphics[width=\columnwidth]{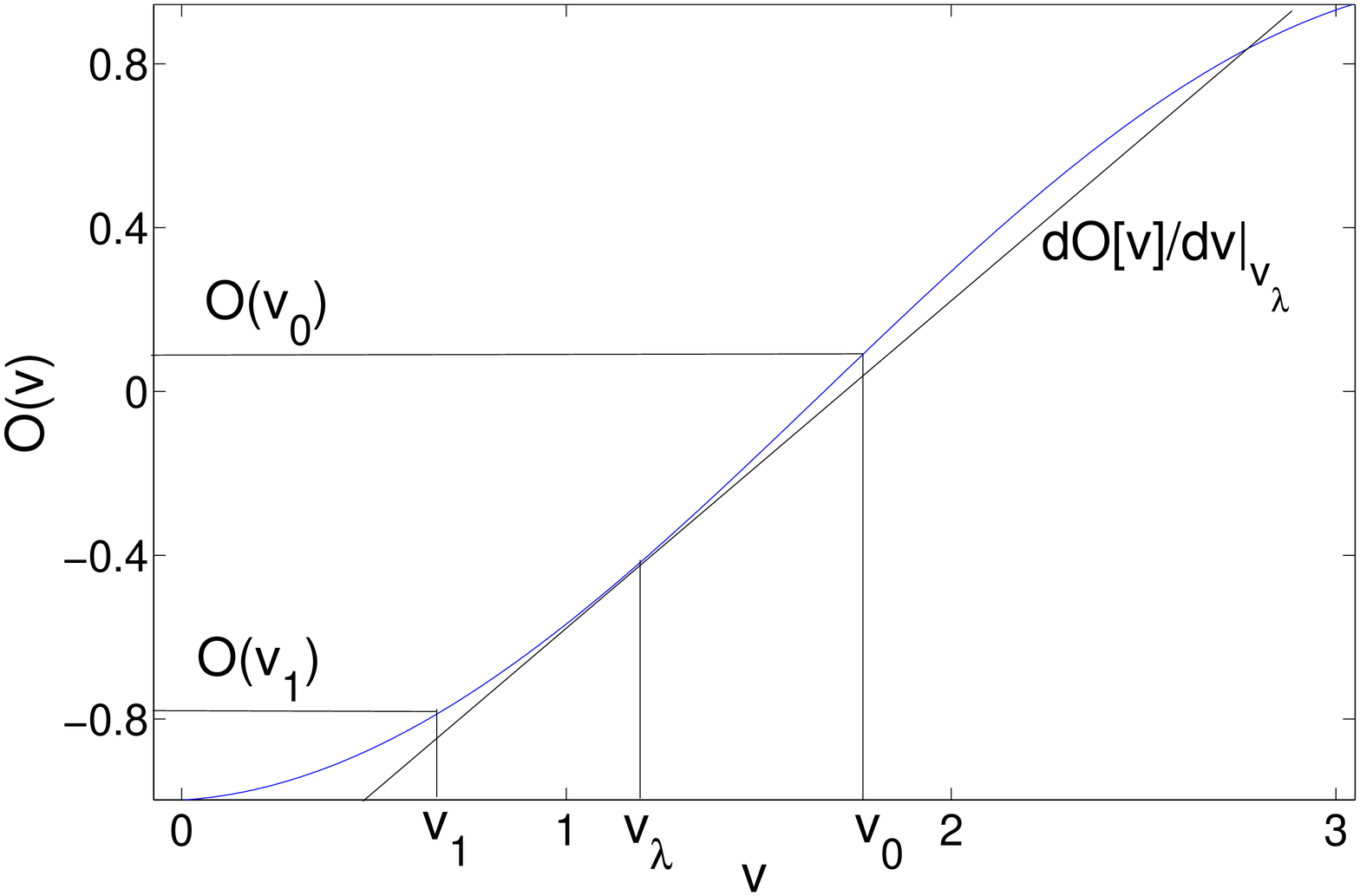} 
\caption{(color online). A simple example illustrating the meaning of the operator norm $C[v_0,v_1]$ which corresponds to the steepest tangent on the interval $[v_0,v_1]$. }
\label{fig:ExampleMap}
\end{figure}

In this section we will use inequality (\ref{step3}) to derive uniqueness and existence of a fixed-point. The presented proofs follow the logics of the Banach fixed-point theorem \cite{Griffel}.
\\
\\
First we will show uniqueness of a given fixed-point. 
Let us define by $\mathfrak{V}$ the set of potentials with a given set of boundary conditions.
Let us take a potential $v$ out of this set and calculate $n([v], xt)$.
This is then, by definition, a $v$-representable density.
Let us now assume that there is a second fixed-point $u$ in $\mathfrak{V}$, i.e.\ a potential
yielding the same density. Then we find by choosing $\sqrt[p]{\alpha}=2 \,CD$ for this pair of potentials in Eqs.~(\ref{step3}-\ref{step2}) that
\bea
\| v - u \|_{\alpha} = \| \mathcal{F}[v] - \mathcal{F}[u] \|_{\alpha} \leq \frac{1}{2} \| v - u \|_{\alpha}. \nonumber
\eea
This can only be true if 
\begin{eqnarray*}
\| v -u \|_{\alpha} = 0
\end{eqnarray*}
and thus we have $u=v$. 
\\
\\
Let us now address the existence of a solution to Eq.~(\ref{InitialSturmLiouville2}). This is a $v$-representability question for a given density. Before we present the actual proof we give a simplified example that illustrates the physical meaning of the assumptions made for the
response function $\chi$. To do so, we look at an analogy in which potentials are represented by real numbers $v \in [a,b]$ 
and where an observable $O(v)$ is represented as a real function which maps to $\mathbb{R}$. Thus we associate with $v$ a point on the real axis. Following the reasoning of Sec.~\ref{SecLinResp} we then look at the derivative of the function $O$ with respect to some parameter $\lambda \in [0,1]$, where we define $v_{\lambda} = v_0 + \lambda \Delta v$ with $\Delta v = v_1-v_0$. From the fundamental theorem of calculus we can thus derive that
\begin{align*}
O(v_1) - O(v_0) &= \int_{0}^{1} \diff \lambda \frac{\diff O[v_{\lambda}]}{\diff \lambda} \nonumber
\\
&=  \int_{0}^{1} \diff \lambda \frac{\diff O[v_{\lambda}]}{\diff v_{\lambda}}\frac{\diff v_{\lambda}}{\diff \lambda}  =   \chi [v_0,v_1]  \Delta v, \nonumber
\end{align*}
where $\int_{0}^{1} \diff \lambda \,\diff O[v_{\lambda}]/\diff v_{\lambda} = \chi[v_0,v_1]$. Then we find that
\begin{eqnarray*}
C[v_0,v_1] = \max_{v \in [v_0,v_1]} \left\lvert \frac{\diff O[v]}{\diff v}  \right\rvert
\end{eqnarray*}
is an upper bound for the slope of the tangent and we find that
\begin{eqnarray*}
\abs{ O(v_1) - O(v_0) } \leq C[v_0,v_1] \abs{\Delta v},
\end{eqnarray*}
in correspondence with Eq.~(\ref{ineq1}). Thus if $v_1$ approaches $v_0$ also $O(v_1)$ approaches $O(v_0)$, i.e.\ $O(v)$ is continuous on $[v_0,v_1]$ (see Fig.~(\ref{fig:ExampleMap})). 
If there is a maximum slope or derivative of the function $O(v)$ when we range over all $v$ in $[a,b]$ then there exists 
a constant
\begin{eqnarray*}
C_{\sup} = \sup_{v_0,v_1} C[v_0,v_1],
\end{eqnarray*}
when we range over all pairs $v_0$ and $v_1$. This means physically that small changes in $v$ cannot lead to arbitrarily large changes
in the observable $O(v)$.

We will now show that similar assumptions for the response function (\ref{linresp}) imply the existence of a fixed point for $\mathcal{F}$. We see from Eq.~(\ref{opnorm}) that the constant
$C=C(T)$ in Eq.~(\ref{step1}) is dependent on the response function $\chi$ and hence via Eq.~(\ref{linresp}) on potentials $v_0$ and $v_1$, i.e.\ $C=C[v_0,v_1]$. We assume that a constant $C_{\textrm{sup}}=\sup_{v_0} C \bigl[v_0,\mathcal{F}[v_0]\bigr]$ 
exists when we range over all potentials $v_0$ in the domain $\mathfrak{V}$ of potentials with a certain boundary condition. 
Physically this amounts to assume that one cannot induce arbitrarily strong changes in the internal forces by weakly perturbing the quantum system during a finite time. Following our simplified reasoning from above, we essentially presume that the slope of the tangent of $q[v]$ does not become infinite.

Let then $v_k=\mathcal{F}^k [v_0]$ denote the $k$-fold application of the mapping $\mathcal{F}$ on a given initial potential $v_0$
and choose $\sqrt{\alpha} > \, C_{\textrm{sup}}D$. Then Eq.~(\ref{step3}) with $a=C_{\textrm{sup}} D/\sqrt{\alpha}$ implies $\| v_{k+1} - v_{k} \|_{\alpha} \leq a^k \| v_1 - v_0 \|_{\alpha}$ which means that the $v_k$ are a Cauchy series.
Since the set of potentials is a Banach space with the norm $\norm{ . }_{\alpha}$ \cite{Evans2010} and therefore complete, this series
converges to a unique $v$ , i.e.\ $v_k \rightarrow v$ for $k \rightarrow \infty$. 
According to our assumption the response function of Eq.~(\ref{linresp}) exists and hence $q$ is functionally differentiable and 
consequently continuous as a functional of $v$.
Therefore $\lim_{k \rightarrow \infty} q[v_k] = q[v]$ which means that $v$ solves Eq.~(\ref{InitialSturmLiouville2})
and hence is a fixed point. This establishes the existence of a Kohn-Sham system corresponding to the density $n$ in Eq.~(\ref{InitialSturmLiouville2}) provided there is a supremum $\sup_{v_0} C \bigl[v_0,\mathcal{F}[v_0] \bigr]$
when we range over potentials $v_0$ in a non-interacting system.

\section{Periodic densities}

\label{SecCurrentDensity}

Let us summarize what we have found so far. The iteration $v_k = \mathcal{F}^{k}[v_0]$ will converge to an external potential giving a certain density $n(x t)$ in an appropriately chosen $\alpha$-norm. The mapping $\mathcal{F}$ from the set of potentials $\mathfrak{V}$ onto itself depends on the chosen density via the solution of Eq.~(\ref{InitialSturmLiouville3}). Thus we have to invert the Sturm-Liouville operator in every iterative step. Only if the problem is regular at the boundary (see Appendix~\ref{AppendixSturmLiouville}) we can invert in a straightforward manner by simply integrating twice. Otherwise we cannot right away fix the boundary conditions for $n v'$ or $v$. For simplicity we now want to restrict our considerations in the following to the regular case.

We will make the assumption of the regular case explicit by assuming that the density is strictly positive, i.e.\ $n(x t) > \epsilon > 0$. This amounts to assume periodic boundary conditions for the quantum system. 
Then we can directly use Eq.~(\ref{vsol2}). It turns out to be convenient to define
\be
\xi (xt) =\int_a^x \diff  y \, \zeta (yt) = \int_a^x \diff  y \, ( q([v_0], yt)- \partial_t^2 n(yt) ).
\label{xi}
\ee
Then by partial integration and using periodicity we find from Eq.~(\ref{vsol2}) that
\begin{align}
v_1 (x t ) &= - \int_a^b \diff y \, \bigl(\partial_y K_t (x,y) \bigr)\,\xi (yt) \nonumber \\
&= - \int_a^x  \diff y \, \frac{\xi (yt)}{n(yt)} + c(t) \int_a^x  \diff y \, \frac{1}{n(yt)},  
\label{v1again}
\end{align}
where we defined the constant
\begin{align}
c(t) = \biggl(\int_a^b \diff y \,\frac{ 1 }{n (yt)}\biggr)^{-1}\!\!\int_a^b \diff y\,  \frac{\xi (yt)}{n(yt)}. \nonumber
\end{align}
In order to eliminate the explicit dependence on $q[v_0]$ in Eq.~(\ref{xi}) we use the
local force equation (\ref{InitialSturmLiouville}) to write
\be
q ([v_0],xt) = \partial_t^2 n([v_0],xt) - \partial_x ( n([v_0],xt) \partial_x v_0 (xt)) . \nonumber
\ee
If we use this in Eq.~(\ref{v1again}) we find 
\begin{multline}
v_1 (xt ) =\int_a^x \frac{\diff y}{n (yt)} \biggr\{ n ([v_0],yt) \partial_y v_0 (yt)  \\
- \int_a^y \diff z \, \partial_t^2 (n ([v_0], zt) - n(zt))\biggr\}  \\
 -  \tilde{c}(t)  \int_a^x   \diff y \,\frac{1}{n (yt)}, 
\label{v1_2}
\end{multline}
where
\begin{multline*}
\tilde{c} (t) = \biggl(\int_a^b \diff y \frac{ 1 }{n (yt)}\biggr)^{-1}\!\!\int_a^b \frac{ \diff y }{n (yt)} \biggl\{
n ([v_0],yt) \partial_y v_0 (yt)  \\
 - \int_a^y \diff z \, \partial_t^2 (n ([v_0],zt) - n(zt))\biggr\}.
\end{multline*}
Note that the form of Eq.~(\ref{v1_2}) is a very convenient one as it only involves densities and potentials, and
there is therefore no need to calculate $q ([v_0],xt)$ explicitly. Furthermore it is clear from the equation that the
constant $\tilde{c}(t)$ makes the potential periodic and that the explicit form fixes the gauge to $v(at)=v(bt)=0$.

\begin{figure}[t]
\includegraphics[width=0.5\textwidth]{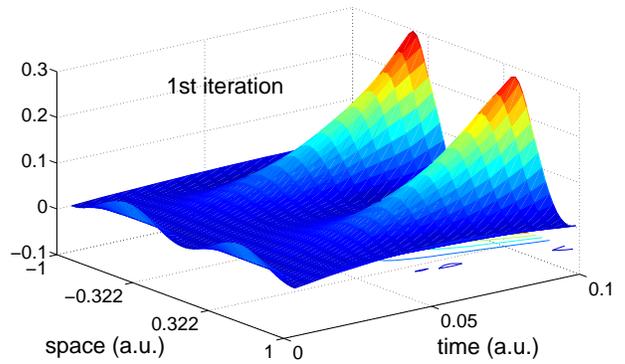} 
\includegraphics[width=0.5\textwidth]{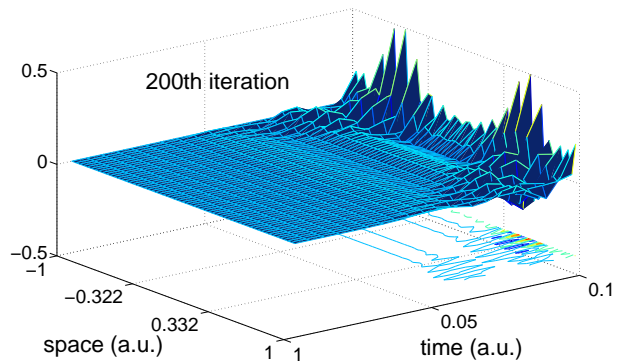} 
\includegraphics[width=0.5\textwidth]{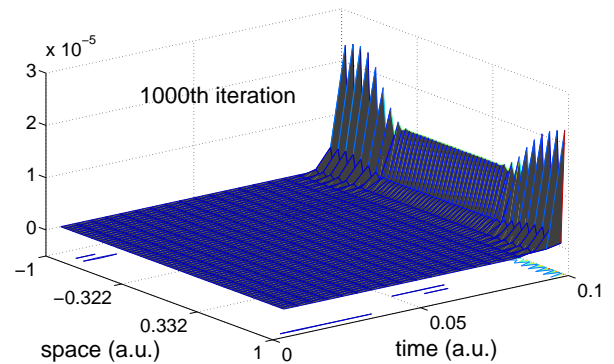} 
\caption{(color online). The difference between the exact potential $v(xt)$ and the iterated potentials $v_1(xt)$, $v_{200}(xt)$ and $v_{1000}(xt)$ in time from $t=0$ until $T=0.1$. Note the change of scale.}
\label{fig:iterpotential}
\end{figure}

Let us now give an explicit example of the iteration scheme.
\begin{figure}[t]
\includegraphics[width=0.5\textwidth]{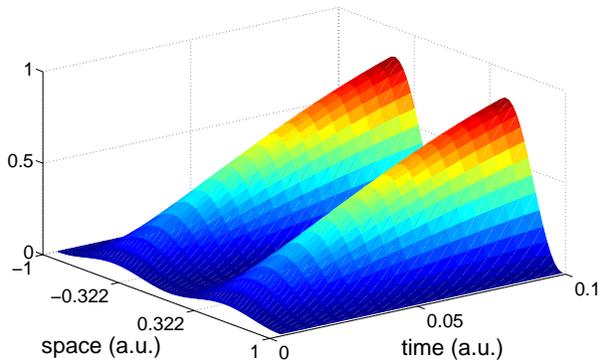} 
\caption{(color online). The potential $v(xt)$ in time from $t=0$ until $T=0.1$.}
\label{fig:potential}
\end{figure}
We take $n(xt)$ to be the density produced in our previous example by potential of Eq.~(\ref{potexample}) (see Fig.~(\ref{fig:potential}))
with the initial state of Eq.~(\ref{initial}). 
Therefore our iteration scheme should recover potential (\ref{potexample}).
We start the iteration with the initial guess $v_0(x t)=0$ in the whole time interval.
As the numerical inaccuracies tend to sum up in time we only look at a small grid and a short time interval. In principle we could perform the iteration for every time step and thus avoid the build up of inaccuracies. However, here we are not so much interested in a long-time propagation but in a proof of principle. Thus we start at $t_0=0$ and go only up to $T=0.1$ atomic units. After one iteration we find (see Fig.~(\ref{fig:iterpotential})) a
first approximation to the exact potential. The approximation becomes worse along the time axis.  After 200 iterations we have almost converged to the exact potential in the first half of the time-interval, while in the second half numerical inaccuracies have build up (see fig.~(\ref{fig:iterpotential})). That the iterative potentials rapidly approximate the exact potential at earlier times is also obvious  from the proof of the fixed-point, as the $\alpha$-norm suppresses differences later in time strongly and thus the convergence is expected to be slower.
As long as these numerical inaccuracies do not go out of hand, for instance by choosing the convergence time interval too large, then we find that after 1000 iterations we have converged to a potential that is practically the same as $v(xt)$ (see Fig.~(\ref{fig:iterpotential})). If we iterate even further we can make both potentials numerically indistinguishable.
This little numerical illustration shows the convergence of the proposed iterative scheme and concludes the presentation of the fixed-point approach to the density-potential mapping in one spatial dimension.

\section{Discussion and Conclusion}

\label{SecCon}

In this work we have given an extensive discussion of the density-potential mappings in quantum dynamics.
We have derived in great detail all necessary equations to rewrite the density-potential mappings as a 
fixed-point question. We note, that a similar global approach was already introduced by Wijewardane and Ullrich 
in \cite{Ullrich}, where they showed the numerical convergence of a time-dependent optimized effective potential 
calculation. The main equations are of Sturm-Liouville type and have been used already in the original work 
by Runge and Gross \cite{RG1984}. In order to uniquely solve these differential equations one needs to pose 
appropriate boundary conditions. These then fix the unique eigensolution to the eigenvalue zero, i.e. 
Eq.~(\ref{eigensolution0}). In the original Runge-Gross proof the potentials were assumed to vanish 
at infinity. If one allows for other boundary conditions then one can find different potentials leading to the same 
density, as has been shown in \cite{Maitra}. Therefore it is obvious that one can prove the Runge-Gross 
theorem only for a set of potentials with common boundary conditions.

Further, we could extend the validity of the original fixed-point proof \cite{FP2011}. While the general 
linear response derivation in Sec.~\ref{SecLinResp} is independent of the dimensionality, we exploit the restriction
to the one-dimensional case when solving the Sturm-Liouville boundary value problems. In the case of periodic densities we can establish uniqueness and existence of a fixed-point for integrable potentials by direct integration.
If the density becomes zero at the boundary we make use of singular Sturm-Liouville theory \cite{Zettl2005}. 
As shown in Appendix~\ref{AppendixContinuum}, if the density goes to zero at the boundary slower or equal 
to a quadratic function we can show uniqueness and existence of a fixed point for square-integrable potentials.

The discussed fixed-point approach to density-potential mappings provides a numerical scheme how to calculate 
the potential for a given density. In Sec.~\ref{SecCurrentDensity} we show that the iterative sequence converges numerically 
to the exact external potential. We point out, that this procedure has several potential applications. For instance, one can calculate 
the exact effective potential of Kohn-Sham TDDFT for a given interacting density. Hitherto this was only possible for special 
cases \cite{KuemmelLein}. Further, one can use the density as the controlling functional variable in quantum control \cite{quantumcontrol}. 

Finally one might be able to extend this fixed-point approach also to other functional theories, e.g. time-dependent current-density-functional theory \cite{Vignale}
, lattice versions of density functional theory \cite{KurthStefanucci,Stefanucci,Verdozzi,LiUllrich} or superconducting systems \cite{Stefanucci}.

\section{Acknowledgments}

M.R. acknowledges  financial support by the Erwin Schr\"odinger Fellowship J 3016-N16 of the FWF (Austrian Science Fonds). K.J.H.G. and R.v.L. acknowledge the Academy of Finland for research funding under Grant No. 127739.

\appendix

\section{Equivalence of the $\alpha$-norms}

\label{AppendixNorms}

In this Appendix we will give a short summary of the functions spaces we are working in. We start by giving the mathematical precise form of the $\alpha$-norm introduced in Sec.~\ref{SecLinResp}: 
\begin{eqnarray*}
\norm{v}_{\alpha}^p = \esssup{t \in [t_0,T]} \left( e^{-\alpha (t-t_0)}\ \norm{ v(t) }^{p} \right).
\end{eqnarray*} 
Here the essential supremum is the supremum up to a set of Lebesgue-measure zero. For any $\alpha \geq 0$ this norm is equivalent to the norm 
\begin{eqnarray*}
\norm{v}_{0}^p = \esssup{t \in [t_0,T]} \norm{ v(t) }^{p} ,
\end{eqnarray*} 
as can be seen by
\begin{eqnarray*}
 e^{-\alpha(T-t_0)} \norm{v}_{0}^p \leq \norm{v}_{\alpha}^p \leq \norm{v}_{0}^p.
\end{eqnarray*}
The space of functions which have finite $\norm{ \cdot }_{0}$-norm is a Banach space, i.e.\ a complete normed vector space, denoted by $L^{\infty}\bigl( [t_0,T],L^p(I)\bigr)$~\cite{Evans2010}. Therefore the Banach space associated with any $\alpha$-norm is isomorphic to this space. Consequently, if a sequence converges in some $\alpha$-norm it converges in every $\alpha$-norm. Thus all $\alpha$-norms are equivalent and we can freely choose the constant $\alpha$ in our calculations. 

Further in the derivation of  inequality (\ref{step1}) we used the operator norm $\tilde{C}(t)$, i.e.\ Eq.~(\ref{opnorm}). The variation therein (up to a normalization) goes over all $g$ for which
\be
\int_{t_0}^{T} \diff t \, \norm{ g(t) }^p < \infty. \nonumber
\ee
Those functions form the Banach space $L^{p}\bigl( [t_0,T],L^p(I)\bigr)$ \cite{Evans2010}. By the simple inequality
\be
\int_{t_0}^{T} \diff t \, \norm{ g(t) }^p \leq \left( T-t_0 \right) \esssup{t \in [t_0,T]} \norm{ v(t) }^{p}.  \nonumber
\ee
We can deduce that $L^{\infty}\bigl( [t_0,T],L^p(I)\bigr) \subset L^{p}\bigl( [t_0,T],L^p(I)\bigr)$ and thus we can include all functions with finite $\alpha$-norm in our considerations.

\section{Sturm-Liouville boundary value problem}

\label{AppendixSturmLiouville}

In what follows we will give a brief sketch of self-adjoint Sturm-Liouville theory following the outline given in reference \cite{Bailey2001}. For a thorough discussion of Sturm-Liouville theory  we refer to \cite{Zettl2005}.

A general Sturm-Liouville boundary value problem reads as
\begin{eqnarray}
\label{MathSturmLiouville}
-\frac{1}{w(x)}\left(\frac{\partial}{\partial x} \left[ n(x) \frac{\partial}{\partial x} \right] + k(x)\right) \varphi(x) = \lambda \, \varphi(x).
\end{eqnarray}
This is an eigenvalue equation for $\varphi$ on an interval $I = {}]a,b[{} \subseteq \mathbb{R}$. In what follows we will use the so-called minimal coefficient conditions
\begin{eqnarray*}
&&n^{-1}, k, w \in L^1_{\text{loc}}(]a,b[), 
\\
&&n(x)>0 \; \text{and} \; w(x)>0 \; \text{on} \;\; ]a,b[ \;\; \text{a.e.}
\end{eqnarray*}
Here ``a.e.'' denotes ``almost everywhere'', i.e.\ up to a set of Lebesque-measure zero, and $L^1_{\text{loc}}(]a,b[):=\{ f: I \rightarrow \mathbb{C} \mid \int_{\alpha}^{\beta} \diff x\, \abs{f(x)} < \infty \quad \forall\; [\alpha,\beta] \subset I  \}$. Note that any continuous function is in $ L^1_{\text{loc}}(]a,b[)$. These conditions are trivially fulfilled in our case, as we have $w(x)=1$ and $k(x)=0$ as well as a density $n(x)$ which in general is continuous. In the following we will restrict ourselves to this special case. With this we will show, that we can always find a self-adjoint realization of the operator $\hat{S} = -\partial_x \left[ n(x) \partial_x \right]$ on the space of square-integrable functions $L^2(I)$. We point out that an operator always consists of a ``rule'', i.e.\ $\hat{S}$, and a ``domain'', i.e.\ which are the functions it is allowed to act on. Actually, depending on $I$ and $n(x)$, we usually have an infinite number of self-adjoint realization which can be distinguished by different (self-adjoint) boundary conditions. We will introduce these boundary conditions in what follows.

From the symmetry condition, i.e.\ $\braket{v}{\hat{S}u} - \braket{\hat{S}v}{u} = 0$, the necessary self-adjoint boundary conditions have to guarantee that
\begin{align}
\label{SymmetryCondition}
\,\,&\!\!\braket{v}{\hat{S}u} - \braket{\hat{S}v}{u} \notag \\
&= \lim_{\beta \rightarrow b^{-}}\bigl\{  u(\beta)\left[  n(\beta) \partial_{\beta} v^{*}(\beta) \right] 
- v^{*}(\beta)\left[  n(\beta)  \partial_{\beta} u(\beta) \right] \bigr\} 
\notag \\
&\hphantom{{}={}} 
- \lim_{\alpha \rightarrow a^{+}}\bigl\{  u(\alpha)\left[  n(\alpha) \partial_{\alpha} v^{*}(\alpha) \right] 
- v^{*}(\alpha)\left[  n(\alpha)  \partial_{\alpha} u(\alpha) \right] \bigr\} \notag \\
& =: \left\{u, v\right\}(b) - \left\{u, v\right\}(a)  = 0
\end{align}
for all $v$ and $u$ in the domain of the operator. Further the domain of any self-adjoint realization is a subset of 
\begin{align*}
\text{D}_{\text{max}} :=  \left\{ v \in L^{2}(I) \left| v, nv' \in \mathcal{AC}_{\text{loc}}(]a,b[), \; \hat{S} v \in L^{2}(I) \right.\right\} ,
\end{align*}
the so-called maximal domain. Here $\mathcal{AC}_{\text{loc}}(]a,b[)$ is the set of locally absolutely continuous functions, i.e.\ $v'$ exists a.e.\ and $v(x) = v(\alpha) + \int_{\alpha}^x \diff y \, v'(y)$ for all subintervals $[\alpha, \beta] \subseteq I$ and $x \in [\alpha, \beta]$. We note here, that the operator $\hat{S}$ becomes positive, i.e.\ $\braket{v}{\hat{S} v} \geq 0$, whenever
\be
 \left. v^{*}(x)\left[  n(x) \partial_{x} v(x) \right] \right|_{a}^{b} = 0  \nonumber
\ee
for all $v$ in its domain. This is a stronger restriction than self-adjointness. Positivity is fulfilled, for instance, if one can choose homogeneous boundary conditions $v(a) = v(b) = 0$. However, one readily sees from the symmetry condition (\ref{SymmetryCondition}) that the boundary conditions will in general not take such a simple form. In fact, the homogeneous boundary condition will in general only lead to a self-adjoint operator for the special case of a so-called regular Sturm-Liouville boundary value problem. In order to differ between the possible cases we introduce the following classification scheme:

The lower endpoint $a$ is called \emph{regular} if
\begin{eqnarray*}
a > -\infty \quad \text{and} \quad \int_a^c \frac{\diff x}{n(x)} < \infty
\end{eqnarray*}
for an arbitrary $c \in I$. The lower endpoint $a$ is called \emph{singular} if either
\begin{eqnarray*}
a = -\infty \quad \text{or} \quad \int_a^c \frac{\diff x}{n(x)}  = \infty
\end{eqnarray*}
for an arbitrary $c \in I$. If the endpoint is singular one either has a \emph{limit-circle endpoint } if for an arbitrary $\lambda \in \mathbb{C}$ and $c \in  I$ any solution of Eq.~(\ref{MathSturmLiouville}) obeys
\begin{eqnarray*}
\int_a^c \diff x \,|\Psi(x)|^2 < \infty,
\end{eqnarray*}
or one has a \emph{limit-point endpoint} if for an arbitrary $\lambda \in \mathbb{C}$ and $c \in  I$ at least one solution of Eq.~(\ref{MathSturmLiouville}) obeys
\begin{eqnarray*}
\int_a^c \diff x  \,|\Psi(x)|^2 = \infty.
\end{eqnarray*}
Keep in mind that the equation is a second order differential equation and thus has two linearly independent solutions to every $\lambda \in \mathbb{C}$. In a similar manner we have a classification scheme for the upper endpoint $b$. An example of two limit-circle endpoints is the well-known Legendre equation, i.e.\ $n(x) = 1-x^2$ on $I ={}]{-1},1[$. Two linearly independent solutions to $\lambda = 0$ are 
\be
\varphi_{0,1}(x) = 1 \nonumber
\ee
and 
\be
\varphi_{0,2}(x) = \int \frac{ \diff x}{n(x)} = \frac{1}{2} \ln\left( \frac{1+x}{1-x} \right). \nonumber
\ee
With this it is easy to check the classification scheme. Depending on the classification scheme of both, the lower endpoint $a$ and the upper endpoint $b$ we can pose different boundary conditions in order to have a self-adjoint operator $\hat{S}$.

The self-adjoint boundary conditions are directly related to the condition (\ref{SymmetryCondition}) and guarantee that the operator $\hat{S}$ is symmetric. If the lower endpoint $a$ is regular then one can pose the self-adjoint boundary condition
\begin{eqnarray*}
A_1 v(a) + A_2 n(a)v'(a) = 0 
\end{eqnarray*}
with $A_1^2 + A_2^2 > 0$. If the lower endpoint $a$ is limit-circle then choose a pair $f,g \in \text{D}_{\text{max}}$ with $f,g \in \mathbb{R}$ and $\{f,g\}(a) \neq 0$. Subsequently 
\begin{eqnarray*}
A_1 \{v,f\}(a) + A_2 \{v,g\}(a) = 0
\end{eqnarray*}
with $A_1^2 + A_2^2 > 0$ is a self-adjoint boundary condition. These boundary conditions ensure that the term depending on $a$ in Eq.~(\ref{SymmetryCondition}) vanishes, i.e.\ $\{u,v \}(a) = 0$. We have according conditions for the upper endpoint $b$. If the endpoint $a$ is limit-point then no boundary condition is needed nor allowed, as the normalizability of the functions becomes a necessary and sufficient condition to make the operator self-adjoint. Then for all functions in the maximal domain $\text{D}_{\text{max}}$ the boundary term $\{u,v\}(a) = 0$ vanishes by construction. We note here, that in the case of regular or limit-circle boundary value problems we can also pose coupled boundary conditions, e.g.\ for a regular Sturm-Liouville boundary value problem $v(a)=v(b)$ and $v'(a)=v'(b)$. For an example of limit-circle boundary conditions we again resort to the Legendre equation. A possible pair $(f,g)$ is $(\varphi_{0,1}, \varphi_{0,2})$ from above. If we choose $A_1 = 1$ and $A_2=0$ for the lower endpoint $a=-1$ as well as the upper endpoint $b=1$ we find the usual Legendre polynomials as eigenfunctions.

It becomes evident that if we have two limit-point endpoints, there is only one possible self-adjoint domain for $\hat{S}$, i.e.\ $\text{D}_{\text{max}}$. Otherwise we have different possible choices for the domain of $\hat{S}$. Irrespective of the choice of boundary conditions, our main interest lies in the spectral properties of the self-adjoint operator. Again we can rely on well-known facts from Sturm-Liouville theory. If both endpoints are either regular or limit-circle then we know that we have a pure point spectrum, i.e.\ only eigenvalues. And for separated boundary conditions we also know that we have simple eigenvalues, i.e.\ every eigenvalue has only one eigenfunction. In the case of coupled boundary conditions certain eigenvalues might have two eigenfunctions as can be seen from the numerical example in Sec.~\ref{SecSturmLiouville}. Depending on the boundary conditions it might occur that the eigenfunction to the eigenvalue zero is not the constant function. In order to assure $\varphi_0(x) = c$ we choose in the regular case either periodic, i.e.\ $v(a)=v(b)$ and $v'(a)=v'(b)$, or homogeneous, i.e.\ $v(a)=v(b)=0$, boundary conditions in accordance to the boundary conditions of our quantum system. For the limit-circle case we can always pose with the pair $\varphi_{0,1}(x)=c$ and $\varphi_{0,2}(x) = \int  \diff y/n(y)$ the so called Friedrich's boundary condition $A_1=1$ and $A_2=0$, i.e.
\be 
\lim_{\alpha \rightarrow a^{+}} n(\alpha) \partial_{\alpha}v(\alpha) = 0,  \nonumber
\ee
and accordingly for the upper endpoint. It is obvious that $\varphi_{0,2}(x)$ cannot fulfill this condition while $\varphi_{0,1}(x)$ does.
Therefore, for the regular as well as the limit-circle case the derivation in Sec.~\ref{SecSturmLiouville} applies directly as we have a lowest non-zero eigenvalue and we can choose the zero-eigenfunction to be the constant function.

If, however, one endpoint is limit-point (here automatically $\varphi_0(x) = c$) one might have a continuous part in the spectrum. Then we have to make sure that the continuous part is gapped away from zero. In the following Appendix~\ref{ap:ContinuumGap} we will show that the continuum is indeed gapped away from zero in the case for the ground state density of a particle in a box by explicit calculation. Subsequently we will show in Appendix~\ref{AppendixContinuum} that for a finite interval the continuum gap depends on the local behavior of the density near the boundaries. In particular, if the behavior of the density can be described as a power series with a lowest power $p$, the continuum is gapped away from zero for $p \leq 2$ and the gap closes for $p > 2$. How the inversion of the Sturm-Liouville operator with a spectral gap in the continuum can be defined is shown in Appendix~\ref{AppendixInversion}.
If the spectral gap closes we can not directly apply the presented fixed-point approach on the set of square-integrable potentials. It seems reasonable to assume that in such a case the density can only be $v$-representable by a non-square-integrable potential.

\section{Continuum gap for particle in the box ground state density}
\label{ap:ContinuumGap}

In order to get some feeling for the continuum and the onset of the gap, we consider the lowest lying unnormalized density of one particle in a box, i.e.\ $n(x) = \cos^2(x)$ on $I = {}]{-\frac{\pi}{2}},\frac{\pi}{2}[$. Then from
\begin{eqnarray*}
\int_{-\frac{\pi}{2}}^{c} \diff x \, \frac{1}{\cos^2(x)} 
= \tan(x) \bigr\rvert_{-1}^{c} \rightarrow \infty
\end{eqnarray*}
we deduce that $a=-\frac{\pi}{2}$ is a singular endpoint.  An according calculation for the upper endpoint reveals that $b=\frac{\pi}{2}$ is also singular. In order to check whether we have a limit-point or a limit-circle endpoint we need two linearly independent solutions to some eigenvalue $\lambda \in \mathbb{C}$. Such two linearly independent solutions to the eigenvalue $\lambda = 0$ in this case are
\begin{align}
\label{eq:0eigenvalueFuncs}
\begin{split}
\varphi_{0,1}(x) &= 1, \\
\varphi_{0,2}(x) &= \int\!\frac{\diff x}{n(x)} = \tan(x).
\end{split}
\end{align}
We can then readily check from the classification scheme that due to
\begin{align*}
\int_{-1}^{c} \diff x \, \tan^2(x)
=  \bigl[ \tan(x) - x \bigr]_{-1}^{c} \rightarrow \infty 
\end{align*} 
the lower endpoint $a=-\frac{\pi}{2}$ is limit-point. And from an according calculation we find that $b=\frac{\pi}{2}$ is a limit-point endpoint as well. Therefore we find that $\text{D}_{\text{max}}$ constitutes the self-adjoint domain.

Now we try to find the general solution for the differential equation of the Sturm--Liouville problem
\begin{align*}
-\frac{\partial}{\partial x} \left[ \cos^2(x) \frac{\partial}{\partial x} \right]\varphi(x)
= \lambda\, \varphi(x).
\end{align*}
First we try to eliminate the $\cos^2(x)$ term by the following coordinate transformation $y = \cos^2(x)$, so $y \in [0,1[$. The differential equation then simplifies to
\begin{align*}
\left(y(1-y)\frac{\partial^2}{\partial y^2} - \left(2y - \frac{3}{2}\right)\frac{\partial}{\partial y} + 
\frac{\lambda}{4y}\right)
h(y) = 0,
\end{align*}
where $h\bigl(\cos^2(x)\bigr) = \varphi(x)$. Note that we will only obtain the solution on half of the interval, since the inverse transformation is $x = \pm\arccos\sqrt{y}$. The solution in the other half of the interval can be reconstructed by demanding continuity of the function and its derivative at $x=0$.

The new differential equation is almost the hypergeometric differential equation, except for the $1/y$ term. This term can simply be dealt with by using the Frobenius method. We write $h(y) = y^pf(y)$ and solve for $p$ such that it annihilates the divergency, which gives $p^{\pm}_{\lambda} = -\frac{1}{4} \pm \frac{1}{4}\sqrt{1- 4\lambda}$. The differential equation for $f(y)$ reduces the following hypergeometric differential equation
\begin{multline*}
y(1-y)f''(y) + \bigl(2p^{\pm}_{\lambda} + \tfrac{3}{2} - 2(p^{\pm}_{\lambda}+1)y\bigr)f'(y) - \\
{} - p^{\pm}_{\lambda}(p^{\pm}_{\lambda} + 1)f(y) = 0,
\end{multline*}
with the solutions expressed in hypergeometric functions $f^{\pm}_{\lambda}(y) = \HPtwoFone\bigl(p^{\pm}_{\lambda}, p^{\pm}_{\lambda} + 1, 2p^{\pm}_{\lambda} + \frac{3}{2}; y\bigr)$, so the full solution becomes
\begin{multline}
\label{eq:eigFuncPM}
\varphi^{\pm}_{\lambda}(x) = \cos^{2p^{\pm}_{\lambda}}(x) \\ 
{} \times \HPtwoFone\bigl(p^{\pm}_{\lambda}, p^{\pm}_{\lambda} + 1, 
2p^{\pm}_{\lambda} + \tfrac{3}{2}; \cos^2(x)\bigr).
\end{multline}
Note that for $\lambda \leq \frac{1}{4}$ the solutions $\varphi^{\pm}_{\lambda}(x)$ are real and for $\lambda > \frac{1}{4}$ they are complex and the plus-minus solutions are each others complex conjugate, ${\varphi^{\pm}_{\lambda}}^*(x) = \varphi^{\mp}_{\lambda}(x)$. One can readily check that the general solution reduces to $\lambda=0$ solutions found before~\eqref{eq:0eigenvalueFuncs}.

Now we found the solutions to the differential equation, we turn back to the Sturm--Liouville problem. First we have to check which solutions to the differential equation are normalizable to separate candidates for the point spectrum and the continuum spectrum. Since the hypergeometric function is bounded over the interval, it is only the possible divergency of $\cos^{2p^{\pm}_{\lambda}}(x)$ at $x=\frac{\pi}{2}$ in~\eqref{eq:eigFuncPM} that can make the norm infinite. The small $x - \frac{\pi}{2}$ behavior of the integrant is given as
\begin{align*}
\abs{\varphi^{\pm}_{\lambda}(x)}^2 \approx \begin{cases}
\bigl(x - \frac{\pi}{2}\bigr)^{-1}					&\text{for $\lambda \geq \frac{1}{4}$} \\
\bigl(x - \frac{\pi}{2}\bigr)^{-1 \pm \sqrt{1-4\lambda}}	&\text{for $\lambda < \frac{1}{4}$},
\end{cases}
\end{align*}
where we used that $\HPtwoFone(a,b,c;0) = 1$. So only the functions $\varphi^+_{\lambda}(x)$ with $\lambda < \frac{1}{4}$ are normalizable and could contribute to the point spectrum.

To determine the point spectrum, we use that the functions should be smooth at $x=0$. Since we effectively only solved the differential equation on half of the interval, say $x \in \bigl[0,\frac{\pi}{2}\bigr[$, this condition is not trivially satisfied. Since we have only one solution per eigenvalue, we can only construct a full solution using $a\,\varphi^+_{\lambda}(x)$ with $a \in \mathbb{C}$ for $x < 0$. Since we have two conditions to satisfy (continuity of the function itself and its derivative), we can only construct a solution if one of these conditions is satisfied automatically, irrespective of the value of $a$. Therefore, either the value or the derivative needs to be zero at $x=0$. The function values and derivatives of the general solutions at $x=0$ can be calculated to be
\begin{align}
\label{eq:phiPM0andDphiPM0}
\begin{split}
\varphi^{\pm}_{\lambda}(0) &= \frac{\sqrt{\pi}\,\Gamma\bigl(2p + \frac{3}{2}\bigr)}
{\Gamma\bigl(p + \frac{1}{2}\bigr)\Gamma\bigl(p + \frac{3}{2}\bigr)}, \\
\frac{\diff\varphi^{\pm}_{\lambda}}{\diff x}(0)
&= -\frac{4p(p+1)}{4p+3}
\frac{\sqrt{\pi}\,\Gamma\bigl(2p + \frac{5}{2}\bigr)}{\Gamma(p+1)\Gamma(p+2)}.
\end{split}
\end{align}
Since $\Gamma\bigl(2p + \frac{3}{2}\bigr) > 0$ for $p \geq -\frac{1}{2}$, we find that $\phi^+_{\lambda}(0) \neq 0$ for all $\lambda \leq \frac{1}{4}$. However, for $p = 0$ and $p = -1$ the derivative vanishes at $x=0$, so we find that $\varphi^+_0(x)$ is the only solution for $\lambda < \frac{1}{4}$, such that the Sturm--Liouville operator is self-adjoint. Therefore, the only eigenfunction is the constant function with eigenvalue zero.

Now we will determine which non-normalizable solutions actually contribute to the continuum spectrum. Although they are not in $\text{D}_{\text{max}}$, they are still required to give a self-adjoint operator. In particular, for $f \in \text{D}_{\text{max}}$ the bracket $\bigl\{\varphi^{\pm}_{\lambda},f\bigr\}\bigl(\frac{\pi}{2}\bigr)$ should vanish. The most divergent function in $\text{D}_{\text{max}}$ we can think of behaves near the boundary as $x^q$, with $q > - \frac{1}{2}$. Working out the bracket, we find
\begin{align*}
\bigl\{\varphi^{\pm}_{\lambda},f\bigr\}\bigl(\tfrac{\pi}{2}\bigr)
&= \bigl(q - 2p^{\pm}_{\lambda}\bigr)\lim_{\beta \to \frac{\pi}{2}^-}
\bigl(\beta - \tfrac{\pi}{2}\bigr)^{2p^{\pm}_{\lambda} + q + 1} \notag \\
&= \bigl(q - 2p^{\pm}_{\lambda}\bigr)\lim_{\beta \to \frac{\pi}{2}^-}
\bigl(\beta - \tfrac{\pi}{2}\bigr)^{\frac{1}{2}\sqrt{1 - 4\lambda} + q + \frac{1}{2}} \notag \\
&= \begin{cases}
0	&\text{for $\lambda \geq \frac{1}{4}$} \\
\infty	&\text{for $\lambda < \frac{1}{4}$},
\end{cases}
\end{align*}
so we find that only functions with $\lambda \geq \frac{1}{4}$ build up the continuum spectrum.

An alternative way to distill the continuum spectrum from the unnormalizable solutions comes from the spectral theorem. The spectral theorem states that the (generalized) eigenfunctions of a self-adjoint operator form a basis for $f \in L^2\bigl(]\frac{\pi}{2},\frac{\pi}{2}[\bigr)$, so in our case
\begin{align*}
f(x) = \tilde{f}_0 + \sum_{s = \pm}\int_{\frac{1}{4}}^{\infty}\!\!\!\diff\lambda \, \tilde{f}_s(\lambda)\varphi^s_{\lambda}(x).
\end{align*}
thus the function $f$ are considered as a wave-packet built from the continuum states $\varphi^s_{\lambda}(x)$ and the constant function. If we would have used the Laplace operator in 1D with $I = {}]{-\infty},\infty[$, the continuum states would have been the plane waves, $\e^{\pm\imagi k x}$, and the integral would already start from 0 and the expansion coefficients $\tilde{f}$ would be the Fourier coefficients of $f$. Further note that the generalized eigenfunction $\varphi^s_{\lambda} \notin L^2\bigl(]\frac{\pi}{2},\frac{\pi}{2}[\bigr)$, so they should be regarded as distributions. Therefore, the functions $\tilde{f}_s(\lambda)$ have to be in the test-function space for the integral to be well defined.

Since we required the generalized eigenfunctions to be such that the Sturm--Liouville operator is self-adjoint, the brackets~\eqref{SymmetryCondition} between the generalized eigenfunctions should also vanish. In particular, for the upper endpoint, $b=\frac{\pi}{2}$, we find in for $\lambda, \lambda' \geq \frac{1}{4}$
\begin{align*}
\bigl\{\varphi^\pm_{\lambda},{}&\varphi^{\pm'}_{\lambda'}\bigr\}\bigl(\tfrac{\pi}{2}\bigr) \notag \\
&= 2\bigl(p^{\pm}_{\lambda} - p^{\mp'}_{\lambda'}\bigr)\lim_{\beta\to\frac{\pi}{2}^-}
\bigl(\beta - \tfrac{\pi}{2}\bigr)^{2p^{\pm}_{\lambda} + 2p^{\mp'}_{\lambda'} + 1} \notag \\
&= 2\bigl(p^{\pm}_{\lambda} - p^{\mp'}_{\lambda'}\bigr)\lim_{\beta\to\frac{\pi}{2}^-}
\bigl(\beta - \tfrac{\pi}{2}\bigr)^{\frac{\imagi}{2}\bigl(\pm\sqrt{4\lambda-1} \mp' \sqrt{4\lambda'-1}\bigr)} \notag \\
&= 0.
\end{align*}
This zero should be considered in a distributional sense, so if the bracket is integrated against test functions, the integral will vanish due to the infinite amount oscillations near the edge, because the generalized eigenfunctions near upper boundary behave as
\begin{align}
\label{eq:genPhiOscillation}
\varphi^{\pm}_{\lambda}(x) \approx \frac{1}{\sqrt{x - \frac{\pi}{2}}}\e^{\pm\frac{\imagi}{2}\sqrt{4\lambda - 1}\ln\left(x - \frac{\pi}{2}\right)}.
\end{align}
The vanishing of the integral can be formulated in a more precise manner by the Riemann--Lebesgue lemma. Similarly also for the lower endpoint we find $\bigl\{\varphi^\pm_{\lambda},\varphi^\pm_{\lambda'}\bigr\}\bigl(-\tfrac{\pi}{2}\bigr) = 0$.

It is now also rather obvious why the solutions $\varphi^-_{\lambda}$ with $\lambda < \frac{1}{4}$ are not generalized eigenfunctions: they do not have the required infinite amount of oscillations to have a vanishing bracket. Indeed, if we check for the upper endpoint the symmetry condition~\eqref{SymmetryCondition} for $\lambda \geq \frac{1}{4}$ and $\lambda' < \frac{1}{4}$ we find
\begin{align*}
\bigl\{\varphi^\pm_{\lambda},{}&\varphi^-_{\lambda'}\bigr\}\bigl(\tfrac{\pi}{2}\bigr) \notag \\
&= 2\bigl(p^{\pm}_{\lambda} - p^-_{\lambda'}\bigr)
\lim_{\beta\to\frac{\pi}{2}^-}
\bigl(\beta - \tfrac{\pi}{2}\bigr)^{2p^{\pm}_{\lambda} + 2p^-_{\lambda'} + 1} \notag \\
&= 2\bigl(p^{\pm}_{\lambda} - p^-_{\lambda'}\bigr)
\lim_{\beta\to\frac{\pi}{2}^-}
\bigl(\beta - \tfrac{\pi}{2}\bigr)^{\pm\frac{i}{2}\sqrt{4\lambda-1} - \frac{1}{2}\sqrt{1-4\lambda'}} \notag \\
&= \infty,
\end{align*}
so indeed we recover that the solutions $\varphi^-_{\lambda}(x)$ with $\lambda < \frac{1}{4}$ do not contribute to the continuum spectrum.

\begin{figure}[t]
\includegraphics[width=\columnwidth]{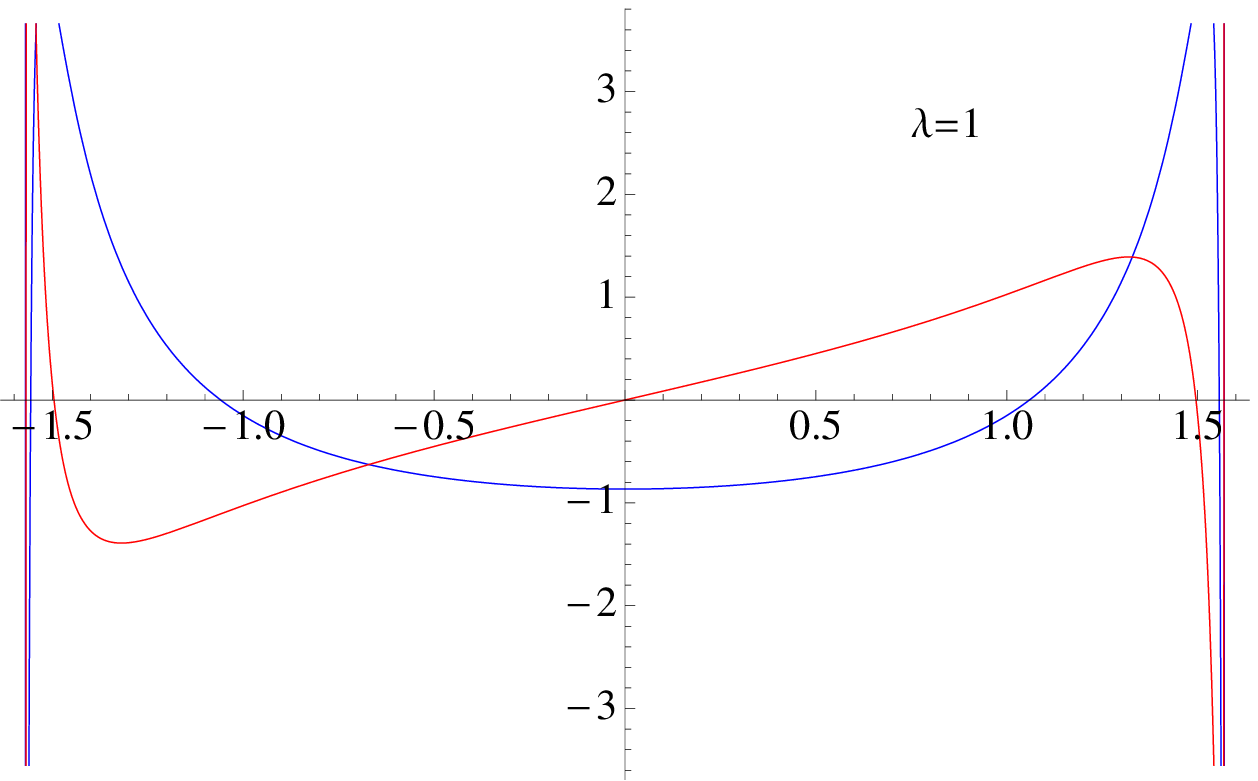}
\includegraphics[width=\columnwidth]{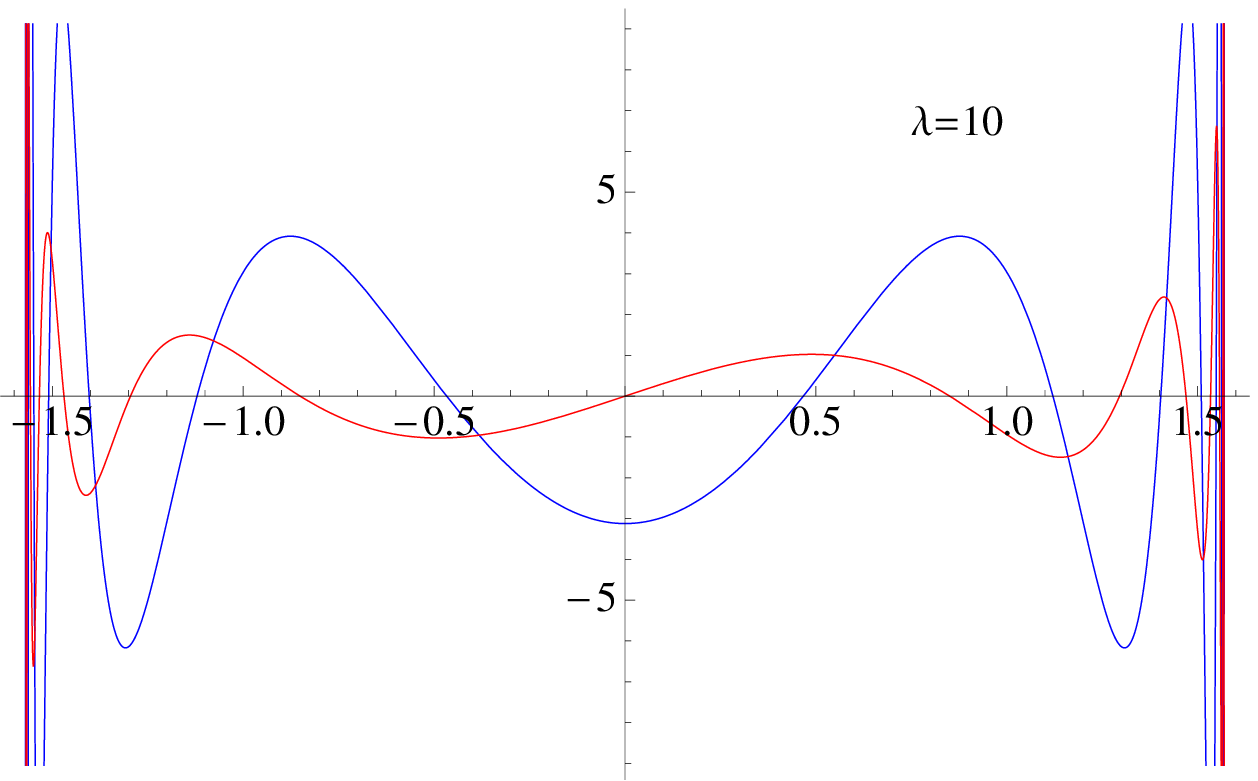}
\includegraphics[width=\columnwidth]{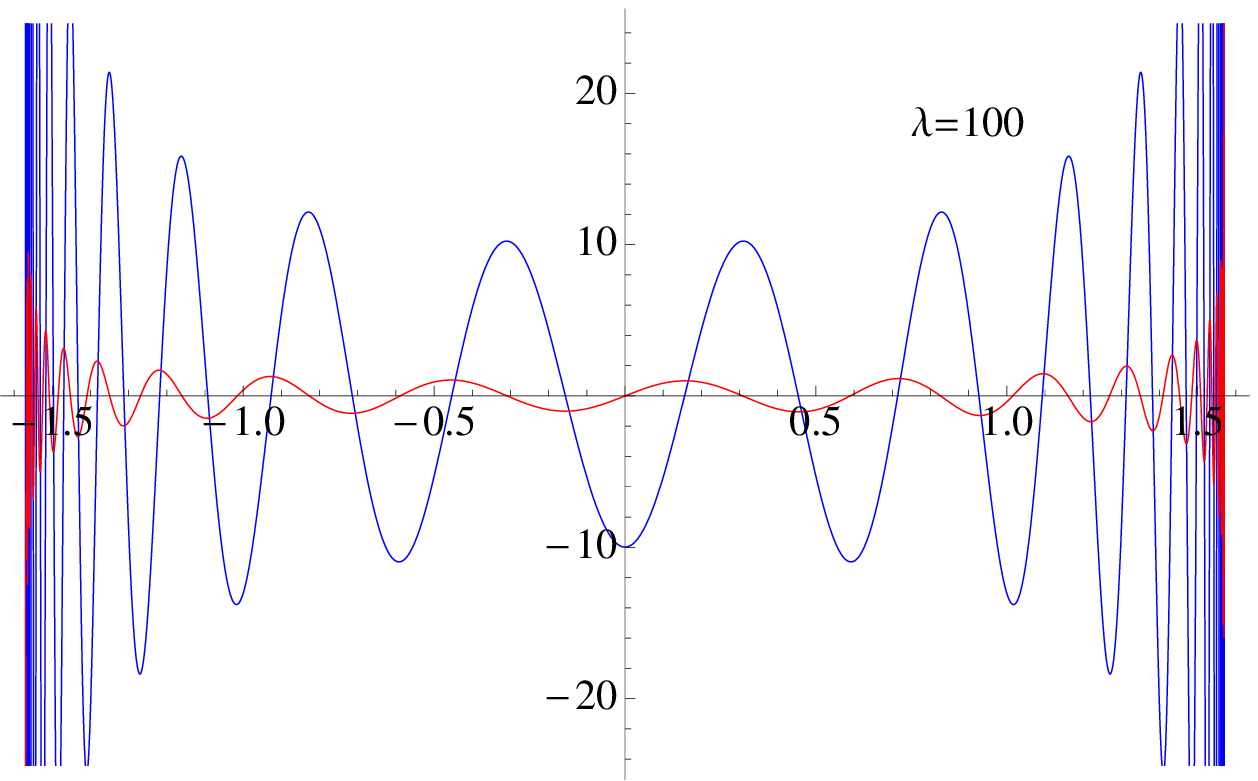}
\caption{(color online). The non-normalized gerade (blue) and ungerade (red) generalized eigenfunctions for $\lambda = 1, 10, 100$.}
\label{fig:contSpectrumEigFuncs}
\end{figure}

The solutions for $\lambda > \frac{1}{4}$ are complex. However, since also the complex conjugate has the same eigenvalue, we can combine them to construct real solutions. In particular we will choose the real solutions to transform as the irreducible representations of the symmetry of the problem, so separate them in gerade, $(\varphi^g_{\lambda})'(0) = 0$, and ungerade, $\varphi^u_{\lambda}(0) = 0$, solutions. With the help of the values at $x=0$~\eqref{eq:phiPM0andDphiPM0}, they readily constructed as
\begin{align*}
\varphi^g_{\lambda}(x) &= \Im(\varphi^+_{\lambda})'(0)\,\Re\varphi^+_{\lambda}(x) \notag \\
&\qquad\qquad\qquad
{} - \Re(\varphi^+_{\lambda})'(0)\,\Im\varphi^+_{\lambda}(x), \\
\varphi^u_{\lambda}(x) &= \sgn(x)\bigl(\Im\varphi^+_{\lambda}(0)\,\Re\varphi^+_{\lambda}(x)\notag \\
&\qquad\qquad\qquad
{} - \Re\varphi^+_{\lambda}(0)\,\Im\varphi^+_{\lambda}(x)\bigr),
\end{align*}
where $\sgn(x)$ is the signum function. These generalized eigenfunction have been plotted in Fig.~\ref{fig:contSpectrumEigFuncs} for the eigenvalues $\lambda = 1, 10, 100$. When approaching the boundaries, the generalized eigenfunctions start to oscillate infinitely fast as mentioned before, due to the logarithmic term in the imaginary exponent~\eqref{eq:genPhiOscillation}. Therefore, the onset of the continuous spectrum is called the oscillation point, $\sigma_0 = \frac{1}{4}$. Thanks to the infinitely many oscillations and the divergence near the edge, the generalized eigenfunctions of the continuous spectrum are able to be orthogonal as required for symmetric operators, although there is an uncountable amount of them. In particular, we have
\begin{align*}
\braket{\varphi^s_{\lambda}}{\varphi^{s'}_{\lambda'}}
= A^s_{\lambda}\delta_{ss'}\delta(\lambda - \lambda'),
\end{align*}
where $s = g, u$ and $A^s_{\lambda}$ is a normalization factor. The precise value of this normalization factor is important if one wishes to compute the resolution of the identity
\begin{align*}
\delta(x-y) =  \frac{1}{\pi} + 
\sum_{s=g,u}\int_{\frac{1}{4}}^{\infty}\!\!\! \diff\lambda \, 
\frac{1}{A^s_{\lambda}}\varphi^s_{\lambda}(x)\varphi^s_{\lambda}(y)
\end{align*}
which is used in Appendix~\ref{AppendixInversion}. However, a spectral gap is all we need to validate our inversion and the associated inequalities.

\section{Continuum on a finite interval}

\label{AppendixContinuum}

The example in the previous section shows that the continuum spectrum is actually a local property if we deal with a finite interval. The infinite amount of oscillations are pilled up near the edges where the density should decay sufficiently fast to zero such that the Sturm--Liouville operator does not blow up, so effectively supporting these oscillations. Assuming that the density decays as $x^p$ near the edges, we can indeed classify how fast the density has to decay to support a continuum and determine its onset.

We will only focus on the lower boundary, since the results immediately carry over to the upper boundary. For convenience, we shift the interval such that the lower boundary is located at $x=0$, so the interval under consideration will be $]0,c[$, where $c$ is some small positive number, such that $n(x) = a\,x^p$ is a good approximation to the real density. Without loss of generality, we assume that the boundary at $c$ is regular and the boundary conditions could be smoothness of the solutions which depend on the precise form of $n(x)$ over the whole interval. First we determine whether the boundary is singular
\begin{align*}
\int_0^c\frac{\diff x}{n(x)} = \int_0^c\diff x\,x^{-p} = \begin{cases}
< \infty	&\text{for $p < 1$} \\
= \infty	&\text{for $p \geq 1$}.
\end{cases}
\end{align*}
Now we need to determine wether the singular cases are limit-circle or limit-point. For that we construct the two solutions at $\lambda=0$
\begin{align*}
\varphi_{0,1}(x) &= 1 \\
\varphi_{0,2}(x) &= \int\frac{\diff x}{n(x)}
= \begin{cases}
\ln(x)			&\text{for $p = 1$} \\
\frac{x^{1-p}}{1-p}	&\text{for $p \neq 1$}.
\end{cases}
\end{align*}
The behaviour of the first solution is not problematic near the boundary for its normalizability. However, the second one might be problematic. In particular we have
\begin{align*}
\int_0^c\!\!\!\diff x\, \ln^2(x) & < \infty \\
\int_0^c\!\!\!\diff x\, \frac{x^{2-2p}}{(1-p)^2}
&= \begin{cases}
< \infty	&\text{for $p < \frac{3}{2}$} \\
= \infty	&\text{for $p \geq \frac{3}{2}$}.
\end{cases}
\end{align*}
The results for the boundary classification have been compiled in Table~\ref{tab:boundaryClassification}.

\begin{table}[t]
\caption{End-point classification}
\label{tab:boundaryClassification}
\begin{ruledtabular}
\begin{tabular}{cll}
$p$ 									&\multicolumn{2}{c}{boundary classification}	\\
\hline
$]{-\infty},1[\vphantom{\big|}$				&regular								\\
$\bigl[1,\frac{3}{2} \bigr[\vphantom{\Big|}$ 	&singular		&limit-circle				\\
$\bigl[\frac{3}{2}, \infty\bigr[$ 				&singular		&limit-point				\\
\end{tabular}
\end{ruledtabular}
\end{table}

To determine whether the singular limit-point cases support a continuum and where it starts, we need to solve the actual differential equation
\begin{align}
-\partial_x\left[ x^p\partial_x\varphi(x)\right] = \lambda\,\varphi(x). \nonumber
\end{align}
The $x^p$ term between the derivatives can be eliminated by the following transformations
\begin{align*}
y &= \ln(x)								&&\text{for $p = 2$},		\\
y &= \frac{2\sqrt{\lambda}}{2-p}x^{\half(2-p)}	&&\text{for $p \neq 2$}.
\end{align*}
Since we have two different coordinate transformations, we need to deal with the cases $p=2$ and $p \neq 2$ separately.

Let us first consider the simplest one, $p=2$. In this case the coordinate transformation turns the non-linear differential equation in a linear one which is straightforwardly solved by standard techniques. Transforming the solutions back, we find for the $p=2$ case the following solutions
\begin{align*}
_2\varphi^{\pm}_{\lambda}(x) = \e^{(-\half \pm \half\sqrt{1 - 4\lambda})\ln(x)}
= x^{-\half \pm \half\sqrt{1 - 4\lambda}}.
\end{align*}
Now we need to determine which solutions might contribute to the point spectrum. Checking the normalization gives
\begin{align*}
\int_0^c\!\!\!\diff x\, \abs{\varphi_{\lambda}^{\pm}(x)}^2
&= \begin{cases}
< \infty	&\text{for $\varphi^+_{\lambda < 1/4}(x)$} \\
= \infty	&\text{otherwise}.
\end{cases}
\end{align*}
Therefore, only the solutions $\varphi^+_{\lambda < 1/4}(x)$ can contribute to the point spectrum. The final selection depends on the boundary conditions at $c$ and therefore, no more can be said  about the point spectrum without additional information. For $\lambda > \frac{1}{4}$ we see that the solutions start to oscillate infinitely fast near the boundary, so we expect $\sigma_0 = \frac{1}{4}$. Indeed, when we work out the bracket~\eqref{SymmetryCondition} with $x^q$ and $q < \half$, we find
\begin{align*}
\bigl\{\varphi^{\pm}_{\lambda},x^q\bigr\}(0)
&= \lim_{x \to 0^+}\bigl({\scriptstyle q - \half \mp\half\sqrt{1-4\lambda}}\bigr)
x^{q + \half \pm\half\sqrt{1-4\lambda}} \notag \\
&= \begin{cases}
0		&\text{for $\bigl(\pm,\lambda \geq \frac{1}{4}\bigr)$} \\
\infty		&\text{for $\bigl(-,\lambda < \frac{1}{4}\bigr)$}.
\end{cases}
\end{align*}
So for a density decaying as $n(x) = a_2 x^2$ near the boundary, we find that there is a continuum starting at $\sigma_0 = a_2/4$.

Now we turn to the cases $p \neq 2$. In these cases we are not so lucky that the coefficients in the differential equation become simply constants, but turns into
\begin{align*}
h''(y) + \frac{p}{2-p}\frac{1}{y}h'(y) + h(y) = 0,
\end{align*}
where we defined $h\bigl(y(x)\bigr) = \varphi(x)$. Note the similarity with Bessel's differential equation; only the term in front of the first derivative is problematic. To eliminate this constant, we write the solution as $h(y) = y^{\alpha}f(y)$ and choose $\alpha$ such that this constant becomes one. Following this strategy, one finds
\begin{align*}
\alpha = \frac{p - 1}{p - 2}
\end{align*}
and the equation for $f(y)$ indeed reduces to Bessel's differential equation
\begin{align*}
y^2f''(y) + yf'(y) + \bigl(y^2 - \alpha^2\bigr)f(y) = 0.
\end{align*}
Performing all the back-substitutions, we can express the general solutions for $p \neq 2$ as
\begin{align*}
_p\varphi^1_{\lambda}(x)
&= x^{\half(1-p)}
J_{\babs{\frac{1-p}{2-p}}}\biggl(\frac{2\sqrt{\lambda}}{\abs{p-2}}x^{\half(2-p)}\biggr), \\
_p\varphi^2_{\lambda}(x)
&= x^{\half(1-p)}
Y_{\babs{\frac{1-p}{2-p}}}\biggl(\frac{2\sqrt{\lambda}}{\abs{p-2}}x^{\half(2-p)}\biggr),
\end{align*}
where $J_{\alpha}(y)$ and $Y_{\alpha}(y)$ are the Bessel functions of the first and second kind respectively.

From these solutions we see that for $p > 2$, the Bessel functions start to oscillate infinitely fast when the approach the boundary. Together with the divergence from the pre-factor they could constitute a continuum spectrum. However, these oscillations are absent for $p < 2$, so we expect in these cases no continuum. Indeed, using that for $y \gg \abs{\alpha^2 - 1/4}$ the Bessel functions behave asymptotically as
\begin{align*}
J_{\alpha}(y) \approx \sqrt{\frac{2}{\pi y}}\cos\left(y - \frac{\alpha\pi}{2} - \frac{\pi}{4}\right), \\
Y_{\alpha}(y) \approx \sqrt{\frac{2}{\pi y}}\sin\left(y - \frac{\alpha\pi}{2} - \frac{\pi}{4}\right),
\end{align*}
we find that all solutions for $p > 2$ are not normalizable and that the brackets~\eqref{SymmetryCondition} vanish for $\lambda \geq 0$. In the case of $p < 2$ we need to use the approximation for small argument of the Bessel functions, $0 < y \ll \sqrt{\alpha+1}$, 
\begin{align*}
J_{\alpha}(y) &\approx \frac{1}{\Gamma(\alpha + 1)}\left(\frac{y}{2}\right)^{\alpha}, \\
Y_{\alpha}(y) &\approx \begin{cases}
\frac{2}{\pi}\bigl(\ln(y/2) + \gamma\bigr)				&\text{for $\alpha = 0$} \\
-\frac{\Gamma(\alpha)}{\pi}\bigl(\frac{2}{y}\bigr)^{\alpha}	&\text{for $\alpha > 0$},
\end{cases}
\end{align*}
where $\gamma \simeq  0.5572$ is the Euler--Mascheroni constant. Using these asymptotic forms we find indeed that for $p < \frac{3}{2}$ both solutions are square integrable and for $\frac{3}{2} \leq p < 2$ only the $_p\varphi^1_{\lambda}$ solutions are normalizable which can be used to construct the point-spectrum. Working out the brackets for the $_p\varphi^2_{\lambda}$ solutions we find that they never satisfy the self-adjointness condition~\eqref{SymmetryCondition}, so there is no continuum spectrum if $p < 2$.

\begin{table}[t]
\caption{Combined results of the lower boundary classification and the onset of the continuum (oscillation point), $\sigma_0$, assuming that the density decays as $a_p\,x^p$ towards the boundary.}
\label{tab:continuumGap}
\begin{ruledtabular}
\begin{tabular}{ccl}
$p$								&$\sigma_0$		&point classification \\
\hline
$\bigl]{-\infty},1\bigr[\vphantom{\Big|}$ 	&$\infty$			&regular point \\
$\bigl[1,\frac{3}{2}\bigr[$				&$\infty$			&limit circle \\
$\bigl[\frac{3}{2},2\bigr[\vphantom{\Big|}$	&$\infty$			&limit point \\
2 								&$a_p/4$			&limit point \\
$\bigl]2,\infty\bigr[$					&0				&limit point \\	
\end{tabular}
\end{ruledtabular}
\end{table}

The results are summarized in Table~\ref{tab:continuumGap}. Note that the situation with a finite continuum gap is rather exceptional; it only occurs for $p=2$. However, physically it is a very relevant one, since the density of a particle in a box typically decays quadratically towards the boundary as we saw in Sec.~\ref{ap:ContinuumGap}. Since the results are also valid for the upper boundary, we see that the side where the density decays fastest will determine whether there will be a continuum and the oscillation points, $\sigma_0$. Further note that in the limit point -- limit point case no additional boundary conditions are required\slash{}needed, so we can always make two linearly independent generalized eigenfunctions, i.e.\ the continuum is doubly degenerate. However, in the case of one limit point and one of the other boundary conditions we need to take a particular linear combination, so in that case the continuum will be simple, i.e.\ non-degenerate.

\section{Sturm-Liouville inversion with a continuous spectrum}

\label{AppendixInversion}

Here we perform the inversion of Sec.~\ref{SecSturmLiouville} for a general self-adjoint operator, which might also have a continuous part in its spectrum. We assume that there is a gap between the eigenvalue zero and the rest of the spectrum. From the spectral theorem for self-adjoint operators we know that every self-adjoint operator $\hat{A}$ has a unique spectral representation in terms of its spectral family (resolution of identity) $E^{A}_{u}$ \cite{BB2003}, which is an operator-valued function from $\mathbb{R}$ onto the set of orthogonal projections. In physics one usually writes this resolution of identity in terms of the Dirac notation, i.e.\ $E_u^{A} \equiv \int_{-\infty}^{u} \diff  u' \; \ket{\Psi_{u'}^{A}}  \bra{\Psi_{u'}^{A}}$. Here one can think of the $\ket{\Psi_u^{A}}$ as generalized eigenfunctions to the operator $\hat{A}$. Thus we find
\begin{eqnarray*}
\hat{A} = \int_{\mathbb{R}} u \, \diff E_u^{A} \equiv \int _{\mathbb{R}} \diff u \, u \, \ket{\Psi_{u}^{A}}  \bra{\Psi_{u}^{A}}.
\end{eqnarray*}
If the spectrum only consists of eigenvalues, i.e.\ it is a pure point spectrum, then the integral becomes a sum over the discrete eigenvalues \cite{BB2003}. The resolution of identity obeys 
\begin{eqnarray*}
\int_{\mathbb{R}} \diff u \, \ket{\Psi_{u}^{A}}  \bra{\Psi_{u}^{A}} = \hat{\mathbb{1}},
\end{eqnarray*}
where $\hat{\mathbb{1}}$ is the identity operator on the Hilbert space. In what follows we assume for notational simplicity that the Sturm-Liouville operator $\hat{S}_t= \partial_x[ n(xt) \partial_x]$ is positive, i.e.\ its spectrum is in $[0, \infty[$. The extension to the general case is straightforward. Then we can represent the inhomogeneity of Eq.~(\ref{InitialSturmLiouville3}) as
\begin{align*}
\zeta(t) &= \int_{0}^{\infty} \diff u \, \ket{\Psi_{u}}  \braket{\Psi_{u}}{\zeta(t)} \\
&=  \int_{0}^{\epsilon} \diff u \, \ket{\Psi_{u}}  \braket{\Psi_{u}}{\zeta(t)} 
+ \int_{\epsilon}^{\infty} \diff u \, \ket{\Psi_{u}}  \braket{\Psi_{u}}{\zeta(t)},
\end{align*}
where $\zeta(x t) = q([v_0], xt) -\partial_t n(x t)$ and $\ket{\Psi_u}$ are the generalized eigenfunctions of $\hat{S}_t$. Here we chose $0 < \epsilon < \lambda_1$ where $\lambda_1$ is a lower bound for the non-zero spectrum. By construction we know that $\zeta(t)$ is perpendicular to the $\lambda=0$ eigenspace and thus we have
\begin{eqnarray*}
\zeta(t) =  \int_{\epsilon}^{\infty} \diff u \, \ket{\Psi_{u}}  \braket{\Psi_{u}}{\zeta(t)}
\end{eqnarray*}
Therefore the solution to Eq.~(\ref{InitialSturmLiouville3}) is 
\begin{eqnarray*}
v_1(t) = \int_{\epsilon}^{\infty} \diff u \, \frac{1}{u} \, \ket{\Psi_{u}}  \braket{\Psi_{u}}{\zeta(t)}
\end{eqnarray*}
as can be seen from
\begin{align*}
\hat{S}_t v_1(t) &= \int_{0}^{\infty} \diff u \, u \,  \ket{\Psi_{u}}  \bra{\Psi_{u}} \int_{\epsilon}^{\infty} \diff u' \, \frac{1}{u'} \, \ket{\Psi_{u'}}  \braket{\Psi_{u'}}{\zeta(t)} \nonumber
\\
&= \int_{\epsilon}^{\infty} \diff u' \, \frac{u'}{u'} \, \ket{\Psi_{u'}}  \braket{\Psi_{u'}}{\zeta(t)} = \zeta(t). \nonumber
\end{align*}
Here we used that $\braket{\Psi_u}{\Psi_{u'}} = \delta(u-u')$. Further we can deduce that $v_1(t) \in L^2$, since
\begin{align*}
\|v_1(t)\|^2 &=  \int_{ \epsilon}^{\infty} \diff u \, \frac{1}{u^2} \,   \braket{\zeta(t)}{\Psi_{u}} \braket{\Psi_{u}} {\zeta(t)}  \nonumber
\\
& \leq \frac{1}{\epsilon^2} \| \zeta(t) \|^2 < \infty.
\end{align*}
In a similar manner we can then find
\begin{eqnarray*}
\| v_2(t) -v_1(t) \|^2 \leq \frac{1}{\epsilon^2} \| q([v_1],t) - q([v_0],t) \|^2.
\end{eqnarray*}
If we then take $D^2 = 1/\epsilon^2$ we can derive inequality (\ref{step2}).


\begin{thebibliography}{99}

\bibitem{GF} A.L.\ Fetter and J.D.\ Walecka, \textit{Quantum Theory of Many-Particle Systems} (Dover Publications, 2003).

\bibitem{rDFMT} M.\ Bonitz, \textit{Quantum Kinetic Theory} (Teubner-Verlag, 1998).

\bibitem{Klaas} K.J.H.\ Giesbertz, E.J.\ Baerends and O.V.\ Gritsenko, \href{http://prl.aps.org/abstract/PRL/v101/i3/e033004}{Phys. Rev. Lett. {\bf 101}, 033004 (2008)}.
      
\bibitem{DFT} R.M.\ Dreizler and E.K.U.\ Gross, \textit{Density Functional Theory - An Approach to the Quantum Many-Body Problem} (Springer-Verlag, 1990).

\bibitem{DFT2} E.\ Engel and R.M.\ Dreizler, \textit{Density Functional Theory - An Advanced Course} (Springer-Verlag, 2011).

\bibitem{Peuckert} V.\ Peuckert, \href{http://iopscience.iop.org/0022-3719/11/24/023}{J. Phys. C {\bf 11}, 4945 (1978)}.

\bibitem{RG1984}E.\ Runge and E.K.U.\ Gross, \href{http://prl.aps.org/abstract/PRL/v52/i12/p997_1}{Phys. Rev. Lett. {\bf 52}, 997 (1984)}.

\bibitem{CarstenBook} C. A.\ Ullrich, {\em Time-Dependent Density-Functional Theory} (Oxford University Press, 2012).

\bibitem{GrossKohn} E.K.U.\ Gross and W.\ Kohn, \href{http://www.sciencedirect.com/science/article/pii/S0065327608606000}{Adv. Quant. Chem. {\bf 21}, 255 (1990)}.


\bibitem{RvL1999}R.\ van Leeuwen, \href{http://prl.aps.org/abstract/PRL/v82/i19/p3863_1}{Phys. Rev. Lett.  {\bf 82}, 3863 (1999)}.

\bibitem{MaitraTodorov}    N.T.\  Maitra, T.N.\ Todorov, C.\ Woodward, and K.\ Burke,\href{http://pra.aps.org/abstract/PRA/v81/i4/e042525}{Phys. Rev. A {\bf 81}, 042525 (2010)}. 

\bibitem{RvL2001}  R.\   van Leeuwen, \href{http://www.worldscinet.com/ijmpb/15/1514/S021797920100499X.html}{Int. J. Mod. Phys. B {\bf 15}, 1969 (2001)}.

\bibitem{RPB2010}   M.\  Ruggenthaler, M.\ Penz, D.\ and Bauer, \href{http://pra.aps.org/abstract/PRA/v81/i6/e062108}{ Phys. Rev. A {\bf 81} 062108 (2010)}.

\bibitem{Tokatly2011} I.V.\ Tokatly ,  \href{http://prb.aps.org/abstract/PRB/v83/i3/e035127}{Phys. Rev. B {\bf 83} 035127 (2011)} .

\bibitem{FP2011}
M.\ Ruggenthaler and R.\ van Leeuwen, \href{http://iopscience.iop.org/0295-5075/95/1/13001/}{ Europhys. Lett. {\bf 95}, 13001 (2011)}.

\bibitem{Zettl2005} A.\ Zettl, \textit{Sturm-Liouville Theory} (American Mathematical Society, 2005).

\bibitem{BB2003} Ph.\ Blanchard, E.\ Br\"uning, \textit{Mathematical Methods in Physics} (Birkh\"auser, 2003).

\bibitem{Martin1959} P.C.\ Martin and J.\ Schwinger, \href{http://prola.aps.org/abstract/PR/v115/i6/p1342_1}{Phys. Rev. {\bf 115}, 1342 (1959)}

\bibitem{esssup} To be precise we should take the essential supremum, which is the supremum up to a set 
of measure zero in the chosen function space of potentials. However, for sake of simplicity we leave out this 
mathematical detail at this point. The precise definition is given in Appendix \ref{AppendixNorms}.

\bibitem{Walter} W.\ Walter, \textit{Gew\"ohnliche Differentialgleichungen} (Springer-Verlag, 1996).

\bibitem{Bielecki} A. Bielecki, Bull. Acad. Polon. Sci. IV, 261 (1956). 

\bibitem{Light} W. A. Light, {\em An Introduction to Abstract Analysis}, (Chapman and Hall, 1990).

\bibitem{MP2010} M.\ Ruggenthaler, M.\ Penz, and D.\ Bauer, \href{http://iopscience.iop.org/1751-8121/42/42/425207}{J. Phys. A: Math. Theor. {\bf 42} 425207 (2009)}.

\bibitem{MP2011} M.\ Penz and M.\ Ruggenthaler, \href{http://iopscience.iop.org/1751-8121/44/33/335208/}{J. Phys. A.: Math. Theor. { \bf 44}, 335208 (2011)}.

\bibitem{Greensfunctionnote} We note, that the previously defined Green's function of Eq.~(\ref{vsol2}) obeys the equation $-\partial_x[n(xt)\partial_xK_t(x,y)] = \delta (x-y)$ while $\Gamma_t$ fulfills the somewhat different equation $-\partial_x[n(xt)\partial_x \Gamma_t(x,y)] = \delta (x-y) - \varphi_0(xt) \varphi_{0}^{*}(yt)$.

\bibitem{Bailey2001}  P.B.\ Bailey, W.N.\ Everitt, and A.\ Zettl, \href{http://doi.acm.org/10.1145/383738.383739}{ACM Trans. Math. Software \textbf{27}, 143 (2001)}. 

\bibitem{Griffel} D.H.\ Griffel, \textit{Applied Functional Analysis} (Ellis Horwood Ltd., 1985).

\bibitem{Evans2010} L.C.\ Evans, \textit{Partial Differential Equations} (American Mathematical Society, 2010).

\bibitem{Ullrich} H. O.\ Wijewardane and C. A.\ Ullrich, \href{http://prl.aps.org/abstract/PRL/v100/i5/e056404}{Phys. Rev. Lett. {\bf 100}, 056404 (2008)}.

\bibitem{Maitra} N.T.\ Maitra and K.\ Burke, \href{http://pra.aps.org/abstract/PRA/v63/i4/e042501}{Phys. Rev. A {\bf 63}, 042501 (2001)}. 

\bibitem{KuemmelLein} M.\ Lein and S.\ K\"ummel, \href{http://prl.aps.org/abstract/PRL/v94/i14/e143003}{Phys. Rev. Lett. {\bf 94}, 143003 (2005)}. 

\bibitem{quantumcontrol} A.P.\ Peirce, M.A.\ Dahleh and H.\ Rabitz, \href{http://pra.aps.org/abstract/PRA/v37/i12/p4950_1}{Phys. Rev. A {\bf 37}, 4950 (1988) } .

\bibitem{Vignale} G.\ Vignale, \href{http://prb.aps.org/abstract/PRB/v70/i20/e201102}{Phys. Rev. B {\bf 70}, 201102(R) (2004)}.

\bibitem{KurthStefanucci} S. Kurth and G. Stefanucci, \href{http://www.sciencedirect.com/science/article/pii/S0301010411000346}{Chem. Phys. {\bf 391}, 164 (2011)}.

\bibitem{Stefanucci} G. Stefanucci, E. Perfetto, and M. Cini, \href{http://prb.aps.org/pdf/PRB/v81/i11/e115446}{Phys. Rev. B {\bf 81}, 115446 (2010)}.

\bibitem{Verdozzi} C. Verdozzi, \href{http://prl.aps.org/pdf/PRL/v101/i16/e166401}{Phys. Rev. Lett. {\bf 101}, 166401 (2008)}.

\bibitem{LiUllrich} Y. Li and C. Ullrich, 
\href{http://scitation.aip.org/getpdf/servlet/GetPDFServlet?filetype=pdf&id=JCPSA6000129000004044105000001&idtype=cvips&prog=normal}{J. Chem. Phys. {\bf 129}, 044105 (2008)}.

\end{thebibliography}
\end{document}